\documentclass[aps,twocolumn, superscriptaddress, english, nofootinbib, 10pt]{revtex4}

\usepackage{float}

\usepackage{amssymb, amsmath, bm, dcolumn, epsf, graphicx, latexsym, slashed, simplewick}
\usepackage[utf8]{inputenc}
\usepackage{subfigure}
\usepackage{comment}
\usepackage{graphicx}
\usepackage{hyperref}
\usepackage{enumerate}
\usepackage{lipsum}

\usepackage[usenames,dvipsnames]{xcolor}
\hypersetup{
    colorlinks = true,
    citecolor = {MidnightBlue},
    linkcolor = {BrickRed},
    urlcolor = {BrickRed}
}

\newcommand{\ba}{\begin{eqnarray}}
\newcommand{\ea}{\end{eqnarray}}

\begin{document}

\title{Assessing consistency between CMB temperature and polarization measurements, with application to Planck, ACT and SPT}


\author{Adrien La Posta$^{1}$,
Umberto Natale$^{2}$,
Erminia Calabrese$^{2}$,
Xavier Garrido$^{1}$,
Thibaut Louis$^{1}$
\\ 
$^{1}$\emph{Universit\'e Paris-Saclay, CNRS/IN2P3, IJCLab, 91405 Orsay, France}\\
$^{2}$\emph{School of Physics and Astronomy, Cardiff University, The Parade, Cardiff, Wales CF24 3AA, UK
}}
\begin{abstract}
\emph{Planck}'s Cosmic Microwave Background temperature and polarization observations are the premier dataset for constraining cosmological models. Cosmic variance limited temperature at large and intermediate scales today dominates the constraints; polarization provides additional constraining power and further scrutiny of the models. To complete this picture from \emph{Planck}, ground-based experiments, such as the Atacama Cosmology Telescope (\emph{ACT}) and the South Pole Telescope (\emph{SPT}) continue to add temperature and polarization measurements at small scales, allowing for the extraction of competitive cosmological constraints from the $TE$ and $EE$ power spectra. Matching at the same time all these stringent probes is a key challenge and validation step for any cosmological model. In particular, $\Lambda$CDM requires a tight consistency between the temperature and polarization measurements. In this paper, we present a number of methods to identify and quantify possible inconsistencies between temperature and polarization, we apply them to the latest \emph{Planck}, \emph{ACT} and \emph{SPT} data and find no evidence for a deviation from $\Lambda$CDM. Application of these methods will have increased importance for future, more constraining CMB data. 
\end{abstract}

  \maketitle

\section{Introduction}\label{sec:intro}

Our current, most precise estimates of cosmological parameters have been achieved with observations of Cosmic Microwave Background (CMB) temperature and polarization anisotropies~\cite{Planck2018:cosmo,Aiola2020,dutcher2021measurements,Balkenhol2021}, in particular with large and intermediate angular scales measured by the \emph{Planck} satellite~\cite{P18_like} and data at finer resolution from 
the Atacama Cosmology Telescope (\emph{ACT})~\cite{Choi2020} and the South Pole Telescope (\emph{SPT})~\cite{dutcher2021measurements}. These are de facto state-of-the-art limits on cosmological models and the data have significant power for ruling in or out cosmological scenarios over a wide range of scales. Given the importance for the cosmology landscape, assessing the robustness of the results has become a major endeavor for all collaborations and analyses. The quantity and quality of CMB observations has both allowed and demanded an increasing level of scrutiny of the results and of the assumptions made in deriving them (see, e.g.,~\cite{Planck2015:ps,Addison2016,Huang2018,Louis19,Choi2020,Galli2021,LaPosta2021}). For example, in CMB analyses one can derive cosmological parameters using only a subset of scales or frequencies or probes, making different assumptions in analysis methodology such as using or relaxing priors, applying calibrations in different ways etc., and verify that within the expected statistical uncertainty the estimates are consistent. In particular, over the next few years we expect to reach in polarization the same level of cosmic-variance limit in observations that we now have in temperature, obtaining two almost independent routes to cosmological estimates from different probes with similar constraining power. More accurate and sensitive polarization data are expected from the final \emph{ACT} and \emph{SPT} polarization surveys, from the soon-to-be-deployed Simons Observatory (SO)~\cite{SOoverview}, and from the planned CMB-S4 experiment~\cite{CMB-S4}. For these, the CMB analysis framework will need to continue to develop and build tools to validate the results. 

In this paper we present a series of methods to look at consistency -- or to identify inconsistencies -- between CMB temperature and polarization measurements at intermediate and small angular scales, apply them to the latest \emph{Planck}, \emph{ACT} and \emph{SPT} data marginalizing over residual systematics in a joint fit of cosmological and nuisance parameters, and discuss ways to use and expand this methodology in future analyses. 
In all cases we assume a $\Lambda$CDM model which has been proven to be the best-fitting model for these data~\cite{Planck2018:cosmo,Aiola2020,dutcher2021measurements}. 

We summarise the data used and the cosmological framework for our work in Sec.~\ref{sec:data}, present conditional probability method and results in Sec.~\ref{sec:cond}, and transfer function method and results in Sec.~\ref{sec:consist}. We discuss and conclude in Sec.~\ref{sec:Conclusion}.

\section{Data, Likelihoods and basic cosmological model}\label{sec:data}
The methods presented here can be generally applied to any CMB dataset covering Gaussianly-distributed anisotropies, i.e., following the \emph{Planck} terminology the `high$-\ell$' scales. In this paper we consider the three CMB experiments that today set the most stringent limits on cosmological parameters, \emph{Planck}, \emph{ACT} and \emph{SPT}. 

\begin{itemize}
    \item {\bf \emph{Planck}}\label{subsec:planck}\\
Our baseline for \emph{Planck} includes the temperature, temperature cross E modes and E-modes power spectra, $TT$, $TE$, and $EE$, and covariances from the \emph{Planck} 2018 legacy release (PR3)~\cite{P18_like}. We use the \textsc{plik\_lite} CMB-only high-$\ell$ likelihood, implemented in \textsc{Cobaya}~\cite{Cobaya,cobayaascii}, to analyse these data. The spectra entering in the computation of this likelihood have been marginalized over foreground and instrumental systematic uncertainties, and include $TT$ measurements in the range $30<\ell<2500$ and $TE$, $EE$ measurements at $30<\ell<2000$.
\item{\bf \emph{ACT}}\label{subsec:act}\\
We use the latest \emph{ACT} data including temperature and polarization from the forth data release, \emph{ACT} DR4~\cite{Choi2020}. Also in this case, the $TT$, $TE$ and $EE$ power spectra have been marginalized over foregrounds and systematic uncertainties and are contained in the publicly available \textsc{pyactlike} likelihood.
The $TE$ and $EE$ power spectra cover multipoles $326<\ell<4325$, while the $TT$ power spectrum spans $576<\ell<4325$.
\item{\bf \emph{SPT}} \label{subsec:spt}\\
For \emph{SPT} we use the most recent power spectra from the 2020 SPT-3G data release~\cite{dutcher2021measurements} which included only $TE$ and $EE$ spanning multipoles $300<\ell<3000$. The \emph{SPT} data were released with a fortran likelihood characterizing the spectra; we present here and use throughout a python version of this likelihood\footnote{Made available at \url{https://github.com/xgarrido/spt_likelihoods}}. We verify that our python implementation leads to the same results as the official \emph{SPT} constraints published in Ref.~\cite{dutcher2021measurements} in Appendix~\ref{app:SPT3G}. These \emph{SPT} power spectra have not been marginalized over foregrounds and therefore this likelihood includes modelling of polarized Galactic dust both for $EE$ and $TE$ and Poisson-distributed point sources in $EE$ for the three frequency channels $95$, $150$ and $220~\mathrm{GHz}$.
\item{\bf \emph{Low-$\ell$}}\label{subsec:lowl}\\
Although not examined and not scrutinized with the methods presented in this paper, we add low-$\ell$ temperature and polarization information in our cosmological fits. When \emph{Planck} is included in the analysis, we also use the \emph{Planck} \textsc{commander} likelihood which models the non-Gaussian range between $2<\ell<30$. To incorporate the low-$\ell$ polarization information in all data combinations (i.e., even when \emph{Planck} is not included) we use a Gaussian prior for the reionization optical depth, $\tau=0.054\pm 0.007$~\citep{Planck2018:cosmo}. This is a proxy that allows us to treat \emph{Planck}, \emph{ACT} and \emph{SPT} consistently. 
\end{itemize}

In all cases we work within the $\Lambda$CDM model which is described by: the angular scale at recombination $\theta_\mathrm{MC}$ (or alternatively the Hubble constant, $H_0$), the amplitude and the scalar spectral index of primordial perturbations $A_s$ and $n_s$, the baryon and cold dark matter densities $\Omega_bh^2$ and $\Omega_ch^2$, and $\tau$. We also carry forward in the fits the additional parameters needed in the \emph{Planck}, \emph{ACT} and \emph{SPT} likelihoods. These include a global calibration amplitude for \emph{Planck}, $A_\mathrm{Planck}$, and an overall polarization efficiency, $y_p$, for \emph{ACT}. The \emph{SPT} likelihood contains 6 foreground parameters modelling point sources for each cross-frequency spectrum $\nu_1\times\nu_2$ ($D^{\mathrm{ps}, \nu_1\times\nu_2}$) and 4 parameters describing polarized galactic dust emissions in $EE$ ($A_\mathrm{d}^{EE}$, $\alpha_\mathrm{d}^{EE}$) and $TE$ ($A_\mathrm{d}^{TE}$, $\alpha_\mathrm{d}^{TE}$) and 7 nuisance parameters including a temperature/polarization map calibration parameter for each frequency band ($T_\mathrm{cal}^\nu$, $E_\mathrm{cal}^\nu$), and the mean lensing convergence $\kappa$ modelling super-sample lensing.

\section{Conditional probabilities}\label{sec:cond}
In 2015 the \emph{Planck} collaboration presented a first assessment of consistency between its temperature and polarization spectra using conditional probabilities~\cite{Planck2015:cosmo}. For that specific data release the conditionals were used to demonstrate that the temperature-only baseline cosmology was predicting a polarization signal in agreement with the \emph{Planck} observed polarization spectra which were not used to estimate cosmological parameters. For the 2018 legacy release the exercise was repeated to test all possible temperature and polarization combinations~\cite{Planck2018:like}. Here we use the same methodology and introduce some extensions to present new comparisons of temperature and polarization from the same experiment or between different experiments.

\subsection{Computing conditionals}
Given the Gaussian nature of the CMB high-$\ell$ anisotropies, we can test whether a given observed spectrum is consistent with other observations by analytically computing conditional probabilities. 
As done by \emph{Planck}, we define the observed power spectrum vector as
\begin{equation}\label{eqn:data_vec}
    \mathbf{C}_\mathrm{obs} \equiv \left(\mathbf{C}_\mathrm{obs}^{TT},\,\mathbf{C}_\mathrm{obs}^{TE}\,,\mathbf{C}_\mathrm{obs}^{EE}\right)^\top\equiv\left(\mathbf{C}^T_\mathrm{obs},\,\mathbf{C}^P_\mathrm{obs}\right)^\top\,,
\end{equation}
where $T$ and $P$ group spectra containing only temperature and at least one polarization channel (e.g., $TE$, $EE$ or $TE$+$EE$), respectively, and $\top$ is a transpose operation. The covariance matrix of this vector can then be arranged in four blocks
\begin{equation}\label{eqn:cov}
    \mathbf{\Sigma} = \left(\begin{matrix}
    \mathbf{\Sigma}_T & \mathbf{\Sigma}_{TP}\\
    \mathbf{\Sigma}^\top_{TP} & \mathbf{\Sigma}_{P}
    \end{matrix}\right)\,.
\end{equation}
With this notation, the likelihood of this dataset is
\begin{equation}\label{eqn:likelihood}
    -2\ln\mathcal{L}(\boldsymbol{\theta}) = \left(\mathbf{C}_\mathrm{obs}-\mathbf{C}(\boldsymbol{\theta})\right)^\top\mathbf{\Sigma}^{-1}\left(\mathbf{C}_\mathrm{obs}-\mathbf{C}(\boldsymbol{\theta})\right)\,,
\end{equation}
where $\boldsymbol{\theta}$ represents the set of cosmological parameter that defines a theoretical model. 

From here we can take several different approaches to compute conditionals depending on which hypothesis we want to test. In fact, for different hypotheses we need to make two assumptions. The first one is making an initial selection of which $\mathbf{C}_\mathrm{obs}$ components we want to test. The second one is making a selection on the model we are comparing the data with. Specifically, we need to pick a $\boldsymbol{\theta}$ vector, $\boldsymbol{\theta}_\mathrm{build}$, and define the initial probability distribution in $\mathbf{C}$-space as
\begin{equation}\label{eqn:clprob}
    -2\ln \mathrm{P}(\mathbf{C}) = \left(\mathbf{C}-\mathbf{C}_\mathrm{th}\right)^\top\mathbf{\Sigma}^{-1}\left(\mathbf{C}-\mathbf{C}_\mathrm{th}\right)\,.
\end{equation} 
The mean of the distribution is the theoretical spectrum defined as
\begin{equation}
    \mathbf{C}_\mathrm{th} = \left(\mathbf{C}^T_\mathrm{th}(\boldsymbol{\theta}_\mathrm{build}),\,\mathbf{C}^P_\mathrm{th}(\boldsymbol{\theta}_\mathrm{build})\right)\,,
\end{equation}
which can then be used in Eq.~\ref{eqn:likelihood} for $\mathbf{C}(\boldsymbol{\theta})$. As we show later this can be linked to a subset of $\mathbf{C}_\mathrm{obs}$, i.e., some spectra that we choose to trust and therefore we use as benchmark or to another experiment.


We break down the general concept into some specific prescriptions below.\\

\paragraph{Conditioning to temperature observations within the same experiment\\}
The first case to consider is the one that \emph{Planck} adopted in 2015~\cite{Planck2015:cosmo} (and reapplied in 2018~\cite{Planck2018:like}): we work within a single experiment, of this we use the temperature observations and the cosmology that they predict as benchmark and we look at how the observed polarization spectra behave compared to those. This is still relevant today because CMB temperature continues to be the probe providing most of the constraining power for cosmology. 

In this scenario the theoretical predictions are inferred considering only the $TT$ part of Eq.~\eqref{eqn:likelihood}. Thus, if we define the $\boldsymbol{\theta}_\mathrm{build}$ vector as the best-fit point from $TT$, $\boldsymbol{\theta}_\mathrm{bf}^T$, the starting probability distribution in $\mathbf{C}$-space becomes a normal distribution with mean given by
\begin{equation}\label{eqn:thory_from_TT}
    \mathbf{C}_\mathrm{th}\bigl|_T = \left(\mathbf{C}^T_\mathrm{th}(\boldsymbol{\theta}^T_\mathrm{bf}),\,\mathbf{C}^P_\mathrm{th}(\boldsymbol{\theta}^T_\mathrm{bf})\right)\,.
\end{equation}

With this in hand we can compute the conditional probability of the polarization decomposition, $\mathbf{C}^{P}_\mathrm{cond}$, given the observed temperature, $\mathbf{C}^T_\mathrm{obs}$. Noting that, in general, a Gaussian distribution can be rewritten as
\begin{align}
&\left(\mathbf{C}-\mathbf{C}_{\mathrm{th}}\right)^{\top}\Sigma^{-1}\left(\mathbf{C}-\mathbf{C}_{\mathrm{th}}\right)=\nonumber\\
&\left(\mathbf{C}^{T}-\mathbf{C}_{\mathrm{th}}^{T}\right)^{\top}\Sigma^{-1}_{T}\left(\mathbf{C}^{T}-\mathbf{C}_{\mathrm{th}}^{T}\right)+ \left(\mathbf{C}^{P}-\mu\right)^{\top}M^{-1}\left(\mathbf{C}^{P}-\mu\right)\,,\label{eqn:rewrite_normal}
\end{align}
we simply derive the mean and covariance of this normal distribution $\mu \equiv \mathbf{C}^{P}_\mathrm{cond}$ and $M\equiv\mathbf{\Sigma}^P_\mathrm{cond}$ as

\begin{align}
    & \mathbf{C}_\mathrm{cond}^P = \mathbf{C}_\mathrm{th}^P(\boldsymbol{\theta}^T_\mathrm{bf})+\mathbf{\Sigma}_{TP}^\top\mathbf{\Sigma}_{T}^{-1}\left(\mathbf{C}^T_\mathrm{obs}-\mathbf{C}^T_\mathrm{th}(\boldsymbol{\theta}^T_\mathrm{bf})\right)\label{eqn:pol_cond_mean}\\
    & \mathbf{\Sigma}^P_\mathrm{cond} = \mathbf{\Sigma}_P - \mathbf{\Sigma}_{TP}^\top\mathbf{\Sigma}_T^{-1}\mathbf{\Sigma}_{TP}\,.\label{eqn:pol_cond_cov}
\end{align}
\vspace{0.03cm}

\paragraph{Conditioning to polarization observations within the same experiment\\}
We can also do the reverse exercise and start by assuming that the polarization observations are the spectra we consider reliable and want to use them to test the temperature. This is particularly relevant for example for ground-based experiments which can be significantly impacted by atmospheric and foreground contamination in temperature and less so in polarization. Additionally, as we collect more data from \emph{ACT} and \emph{SPT} and from future experiments we will soon transition to a scenario where polarization data dominate the cosmological constraints and therefore provide a very powerful channel for assessing consistency and robustness of the results.

For this, we flip the derivations of the previous case. If we rely only on polarization observations, we can use the $TE$+$EE$ best-fit, $\boldsymbol{\theta}^P_{\mathrm{bf}}$, to compute the theoretical prediction
\begin{equation}\label{eqn:theory_from_P}
    \mathbf{C}_\mathrm{th}\bigl|_P = \left(\mathbf{C}^T_\mathrm{th}(\boldsymbol{\theta}^P_\mathrm{bf}),\,\mathbf{C}^P_\mathrm{th}(\boldsymbol{\theta}^P_\mathrm{bf})\right)\,
\end{equation}
and use this as the mean of the starting probability distribution in Eq.~\eqref{eqn:clprob}. We then have everything needed to compute the conditional probability of the temperature decomposition given polarization observations. Following a similar derivation as before the conditionals now are given by a normal distribution with mean and covariance
\begin{align}
    & \mathbf{C}_\mathrm{cond}^T = \mathbf{C}_\mathrm{th}^T(\boldsymbol{\theta}^P_\mathrm{bf})+\mathbf{\Sigma}_{TP}\mathbf{\Sigma}_{P}^{-1}\left(\mathbf{C}^P_\mathrm{obs}-\mathbf{C}^P_\mathrm{th}(\boldsymbol{\theta}^P_\mathrm{bf})\right)\label{eqn:temp_cond_mean}\\
    & \mathbf{\Sigma}^T_\mathrm{cond} = \mathbf{\Sigma}_T - \mathbf{\Sigma}_{TP}\mathbf{\Sigma}_P^{-1}\mathbf{\Sigma}_{TP}^\top\,.\label{eqn:temp_cond_cov}
\end{align}
\vspace{0.02cm}

\paragraph{Conditioning to data from another experiment\\}
When more than one experiment provides observations which bring significant constraining power for cosmological models, as it's the case for \emph{Planck}, \emph{ACT} and \emph{SPT}, it is important to do a number of inter-experiment tests. This can be done at power spectrum or cosmological parameters level~\cite{Choi2020, Aiola2020, dutcher2021measurements}, looking at residuals, but also with conditionals. We can for example build a joint $\mathbf{C}_\mathrm{obs}$ vector bringing in data from two experiments and then work through subsections \emph{a}-\emph{b} to build conditionals between two subsets of data from two different experiments. However, Eq.~\eqref{eqn:cov} for this needs joint covariance matrices between the two experiments which are currently not available and beyond the scope of this paper. We anticipate that this will be possible in future analyses, combining experiments at the likelihood level will soon no longer be possible without accounting for cross-covariances and these will become standard data products.

Nevertheless we can use the second assumption made in building conditionals to compare different experiments. If we assume that \texttt{Exp1} is the experiment we trust and want to use as benchmark, and \texttt{Exp2} is the experiment providing the observations that we want to test, we can use the best-fit point coming from \texttt{Exp1} for $\boldsymbol{\theta}_\mathrm{build}$ and with it define the probability distribution of \texttt{Exp2}. With this approach we are building a distribution with a mean derived from the cosmology measured by \texttt{Exp1} and with the experimental setup described by the covariance of \texttt{Exp2}. This is possible because in the conditional formalism there is no accounting for the uncertainty in the cosmological measurement, i.e., there is no additional covariance coming from the specific choice of $\boldsymbol{\theta}_\mathrm{build}$.
This allows a large number of options and we explore the most interesting ones in the next subsection.

\subsection{Results}
\begin{figure*}[ht!]
\subfigure[\label{fig:plc_EE_cond_plc_TT}]{\includegraphics[width=0.5\textwidth]{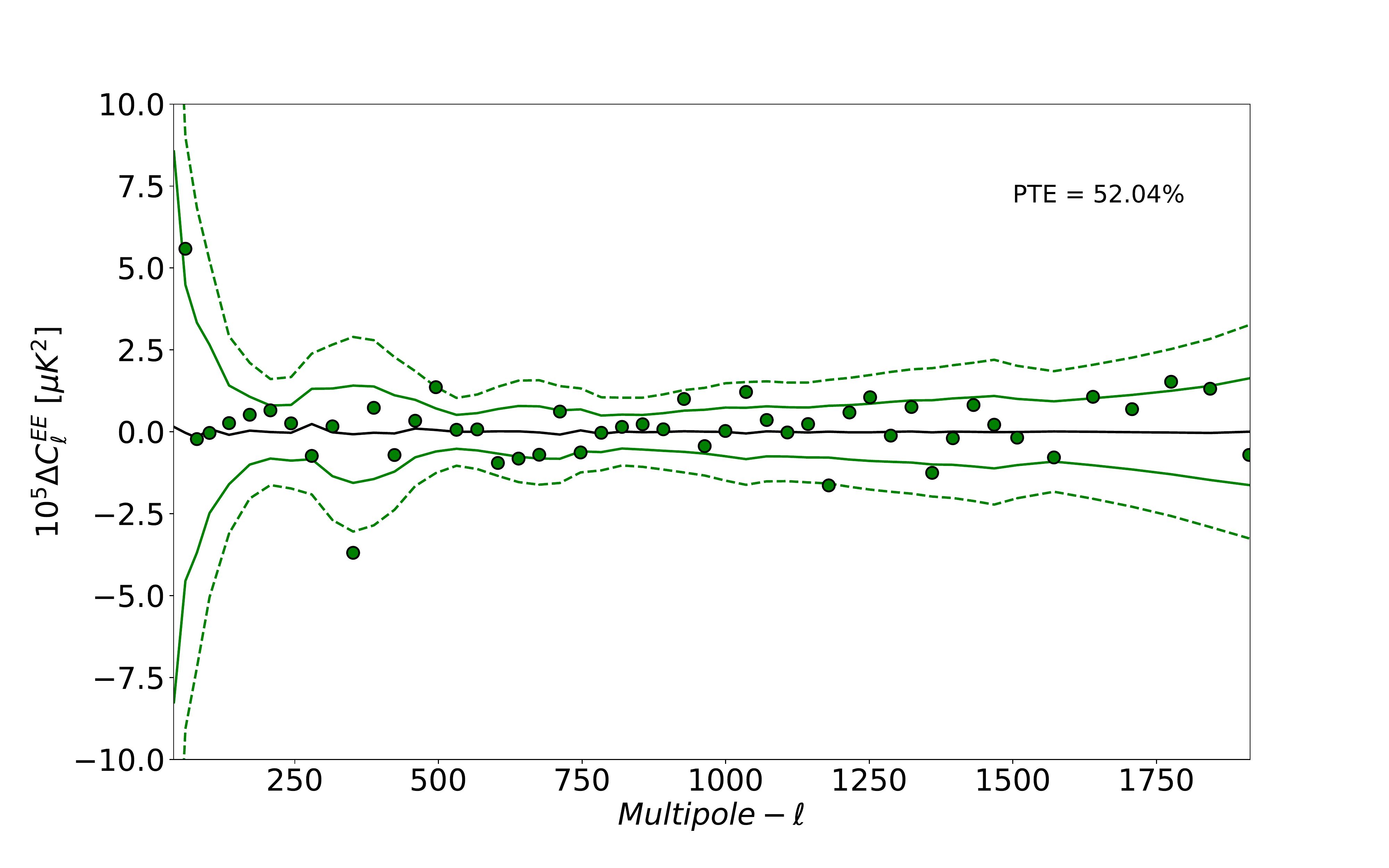}}\hfill
\subfigure[\label{fig:plc_TE_cond_plc_TT}]{\includegraphics[width=0.5\textwidth]{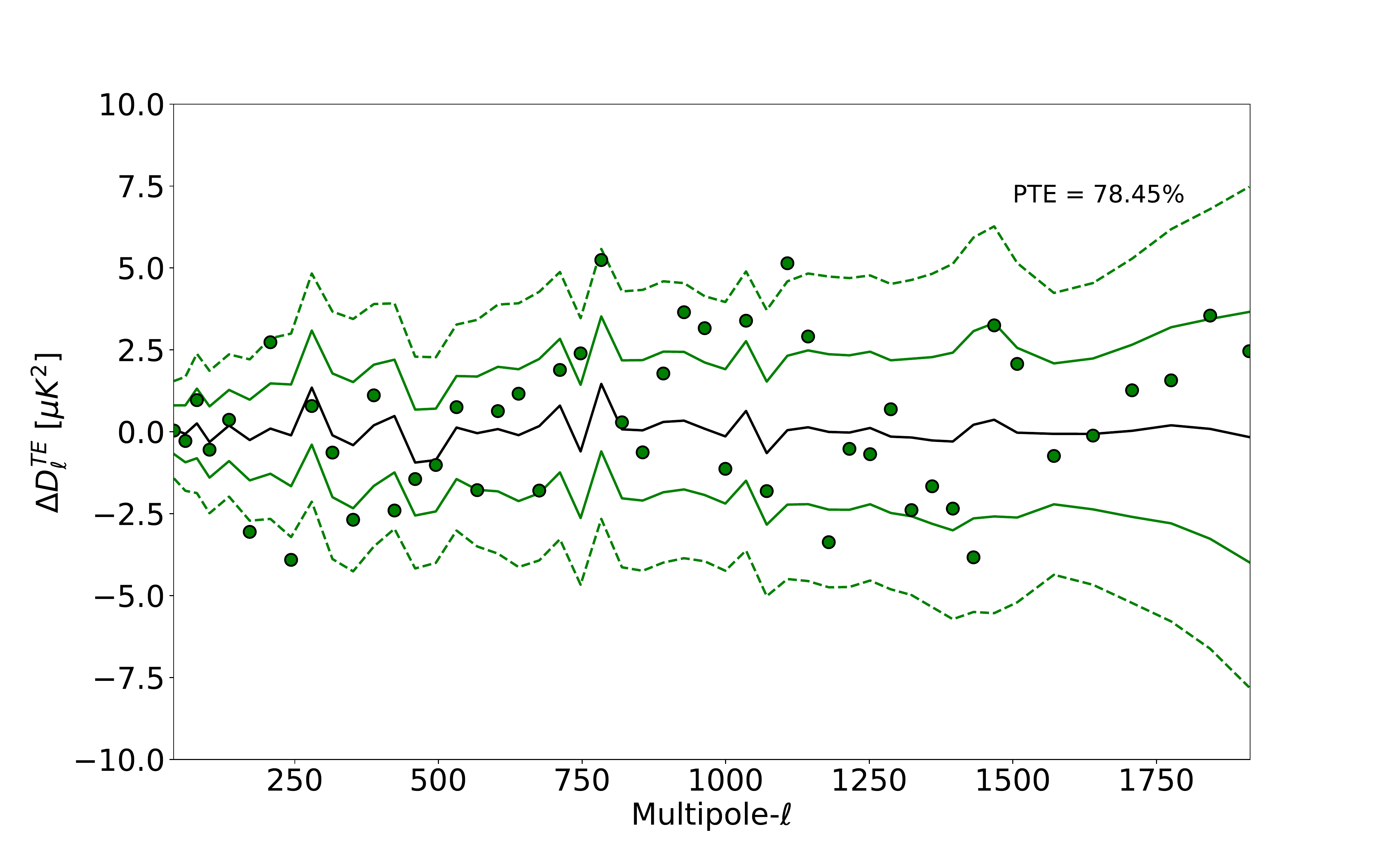}}
\vspace{-0.3cm}
\caption{Conditional probabilities for the \textit{Planck} $EE$ \hyperref[fig:plc_EE_cond_plc_TT]{(a)} and $TE$ \hyperref[fig:plc_TE_cond_plc_TT]{(b)} data given the \emph{Planck} $TT$ observations. The black line represents the difference between the mean of the polarization conditional and the base polarization $\Lambda$CDM best fit determined from the \emph{Planck} TT+low-$\ell$ likelihood. The points are the residuals of the measured bandpowers with respect to the theory. (We bin further the original \textit{Planck} binned likelihood bandpowers reducing the points by a factor of three for visualization purposes.) The colored solid and dashed lines stand for the $1$ and $2\sigma$ uncertainty regions of the conditionals, respectively.}\label{fig:plc_pol_cond_plc_temp}
\end{figure*}
\begin{figure*}[htpb!]
\subfigure[\label{fig:ACT_EE_cond_ACT_TT}]{\includegraphics[width=0.5\textwidth]{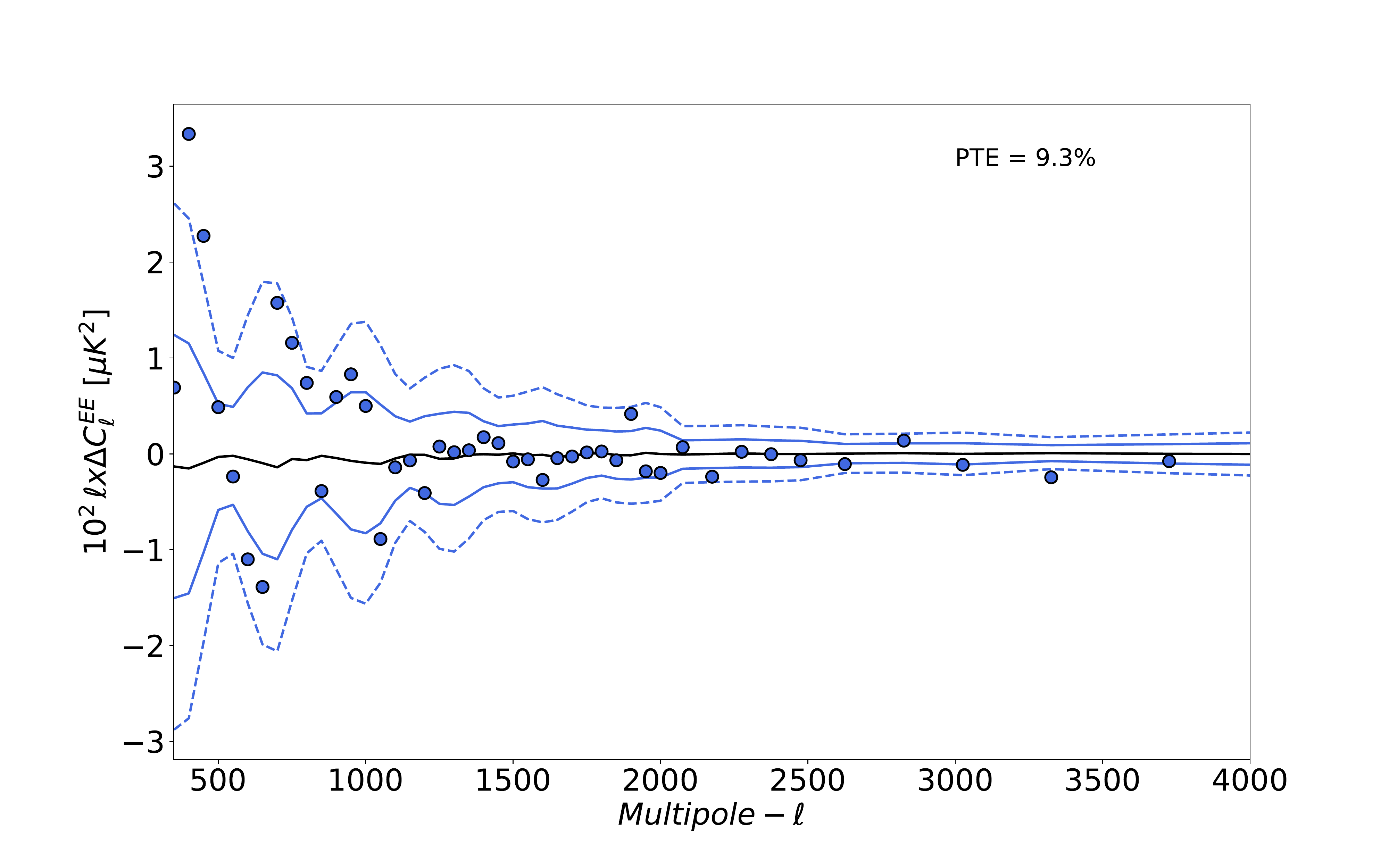}}\hfill
\subfigure[\label{fig:ACT_TE_cond_ACT_TT}]{\includegraphics[width=0.5\textwidth]{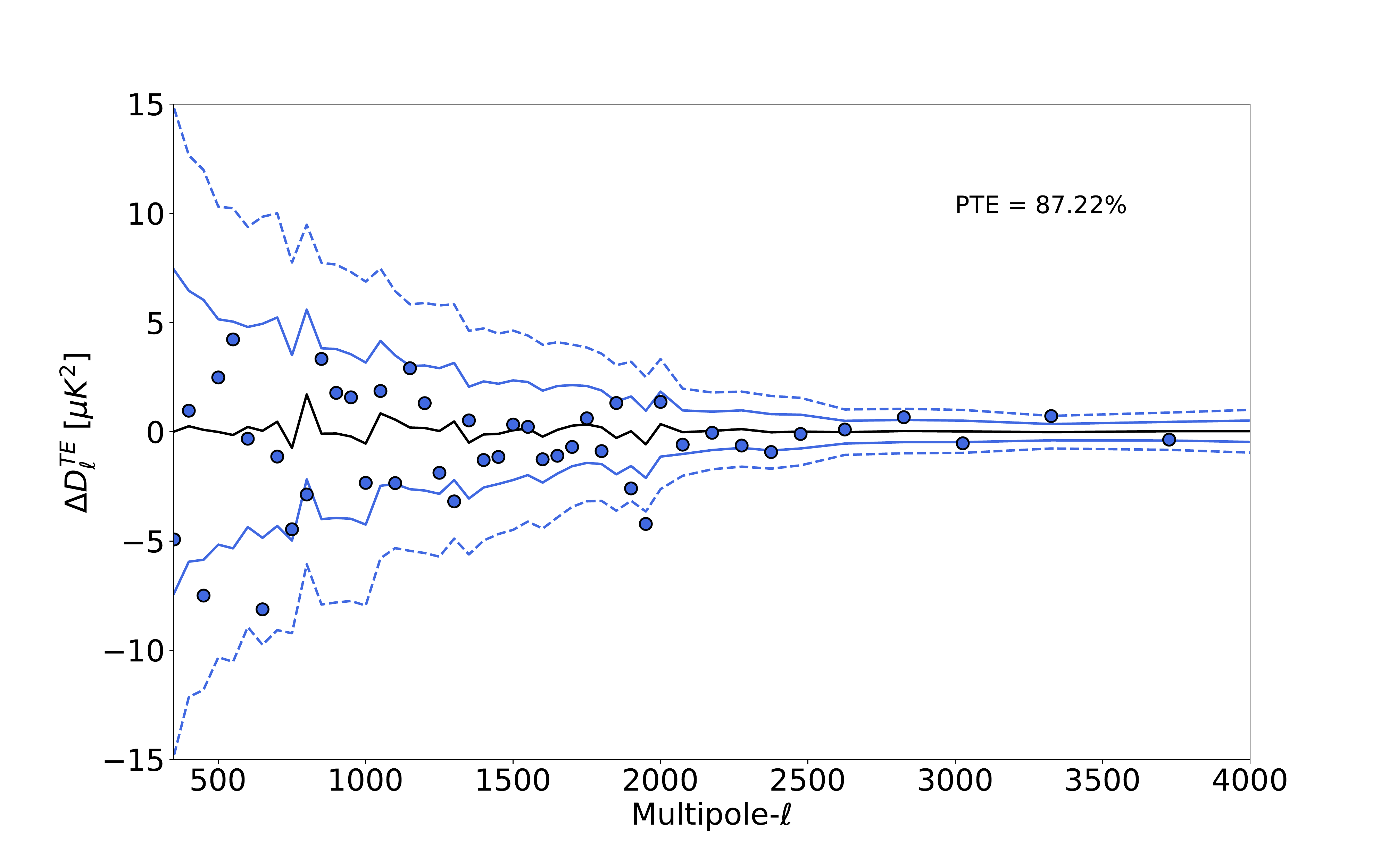}}
\vspace{-0.4cm}
\caption{Conditional probabilities for the \emph{ACT} $EE$ \hyperref[fig:ACT_EE_cond_ACT_TT]{(a)} and $TE$ \hyperref[fig:ACT_TE_cond_ACT_TT]{(b)} data given \emph{ACT} $TT$ observations. The central line shows the difference with the \emph{ACT} $TT$ $\Lambda$CDM best fit. 
}\label{fig:ACT_pol_cond_ACT_temp}
\end{figure*}
\begin{figure}[tp!]
{\includegraphics[width=\columnwidth]{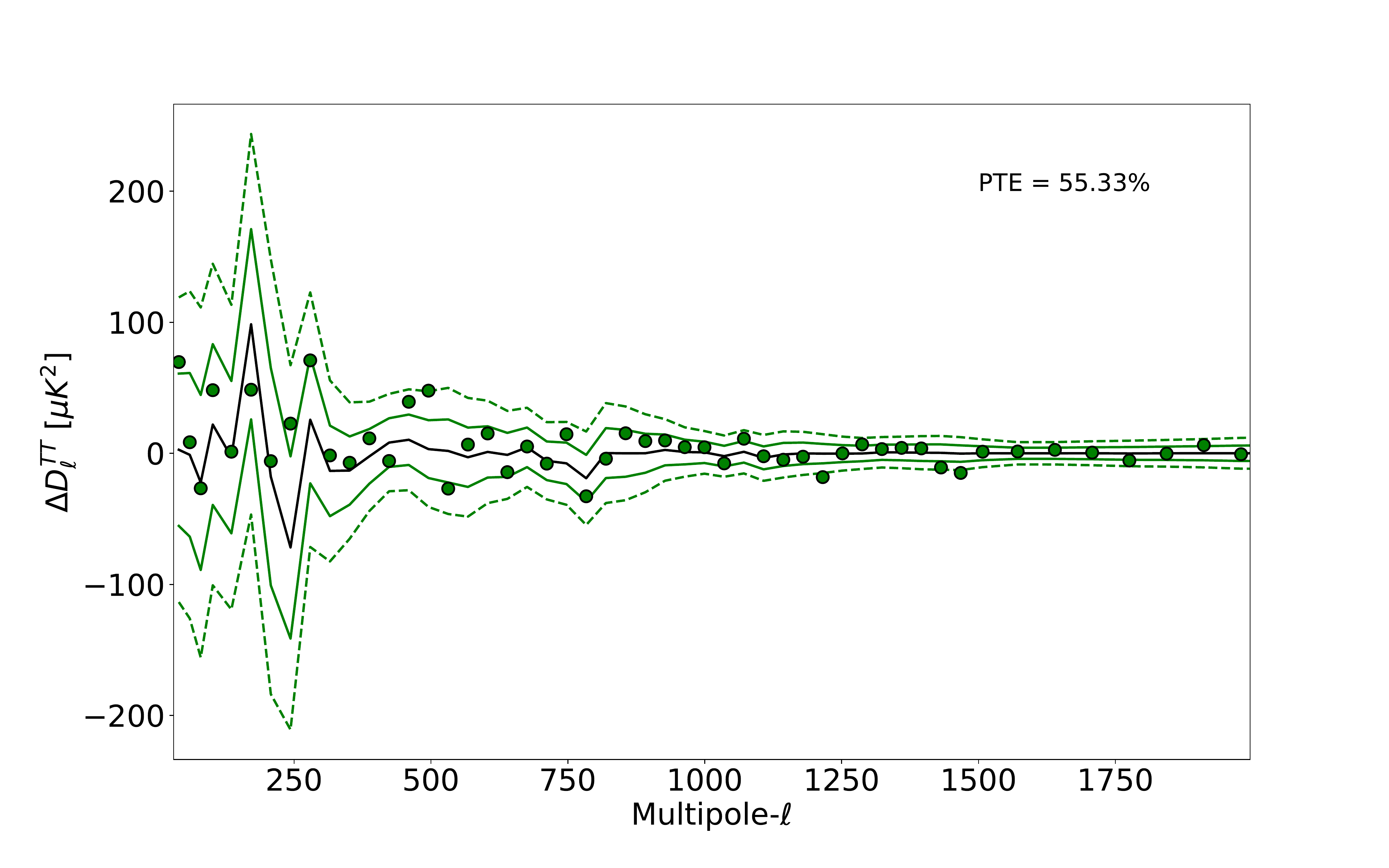}
\vspace{-0.5cm}
\caption{Conditionals for \emph{Planck} $TT$ data given \emph{Planck} $TEEE$ observations. The central line shows the difference with the \emph{Planck} TEEE+low-$\ell$ $\Lambda$CDM best fit.
}\label{fig:plc_TT_cond_plc_TEEE}}
\vspace{-0.3cm}
\end{figure}
\begin{figure}[tp!]
{\includegraphics[width=\columnwidth]{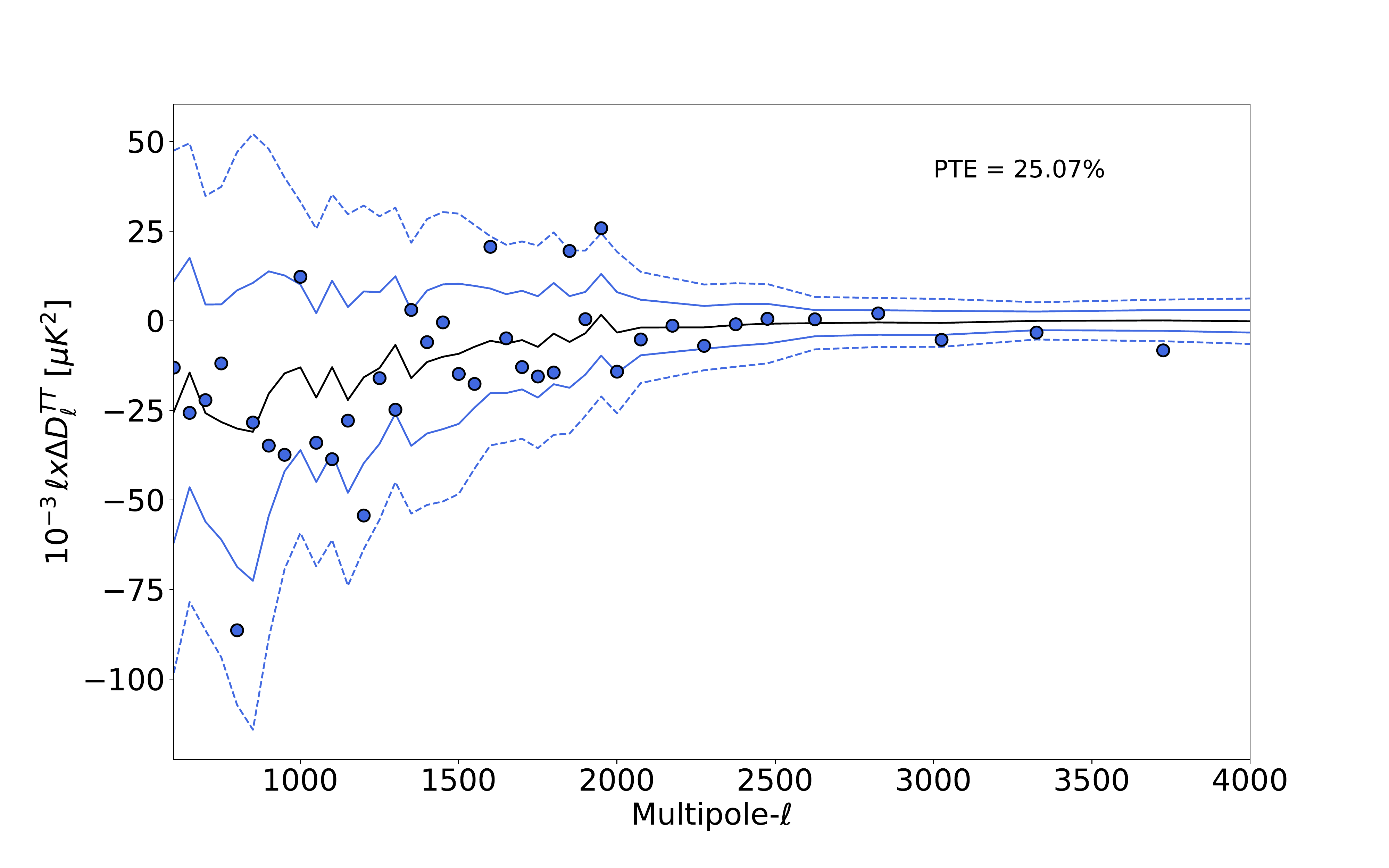}
\vspace{-0.5cm}
\caption{Conditionals for \emph{ACT} $TT$ data given \emph{ACT} $TEEE$ observations. The central line shows the difference with the \emph{ACT} TEEE $\Lambda$CDM best fit. 
}\label{fig:ACT_TT_cond_ACT_TEEE}}
\vspace{-0.3cm}
\end{figure}
\begin{figure*}[htpb!]
\subfigure[\label{fig:ACT_EE_cond_plc_TT}]{\includegraphics[width=0.5\textwidth]{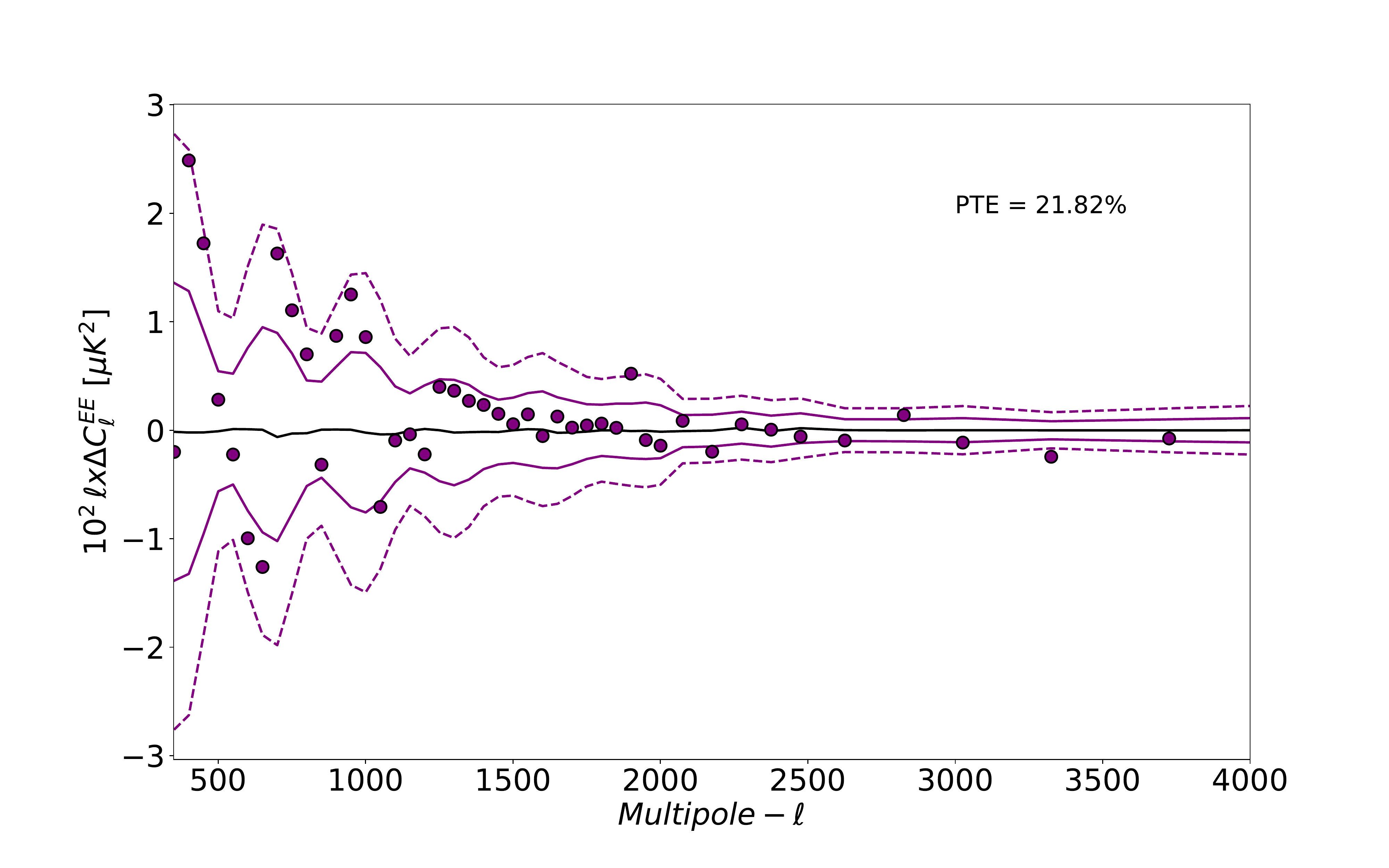}}\hfill
\subfigure[\label{fig:ACT_TE_cond_plc_TT}]{\includegraphics[width=0.5\textwidth]{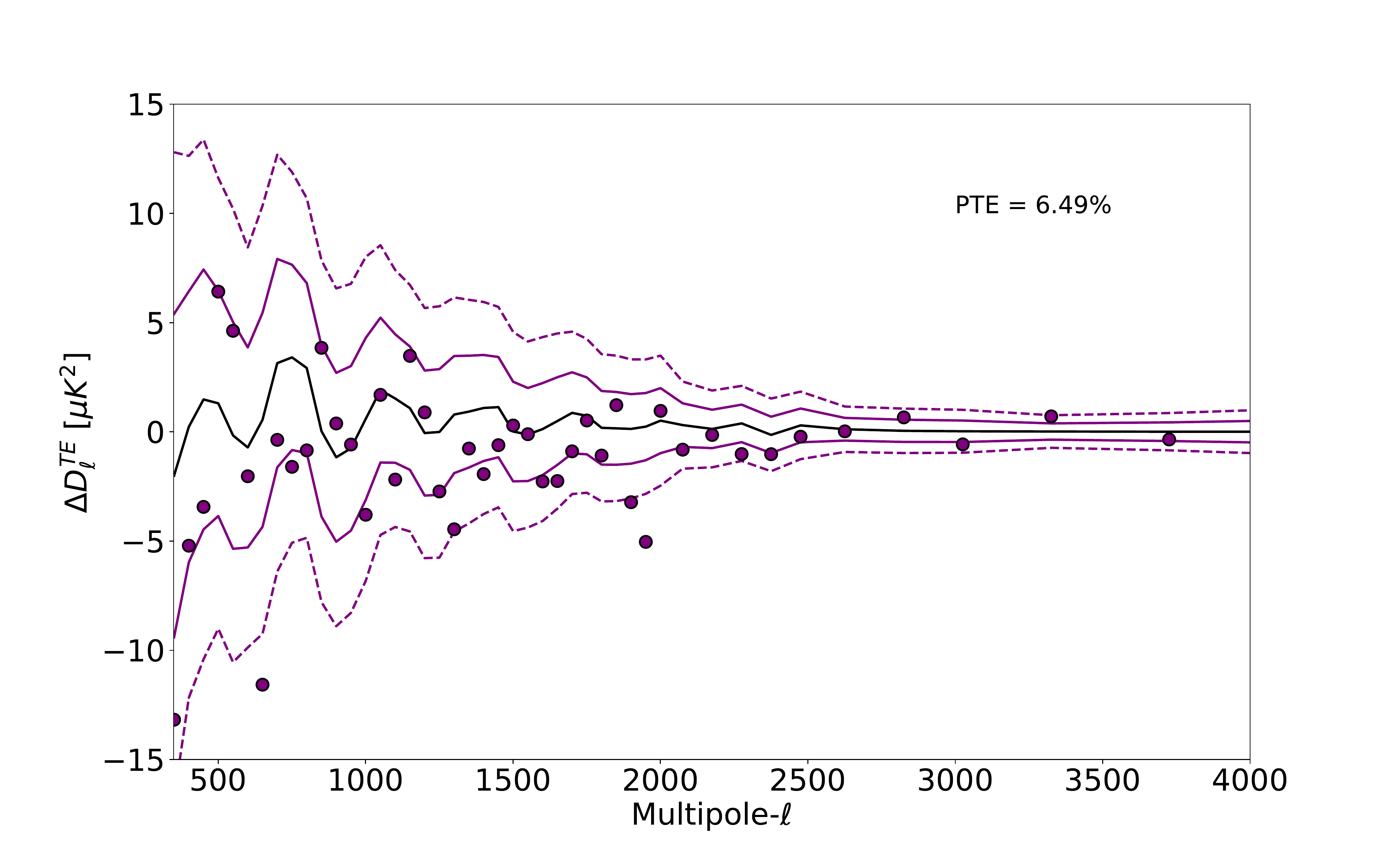}}
\vspace{-0.4cm}
\caption{Conditionals for \emph{ACT} $EE$ \hyperref[fig:ACT_EE_cond_plc_TT]{(a)} and $TE$ \hyperref[fig:ACT_TE_cond_plc_TT]{(b)} data given \textit{Planck} $TT$ observations. The central line shows the difference with the \emph{Planck} TT $\Lambda$CDM best fit.
}\label{fig:ACT_pol_cond_plc_temp}
\vspace{-0.2cm}
\end{figure*}
\begin{figure}[htpb!]
\vspace{-0.4cm}
{\includegraphics[width=\columnwidth]{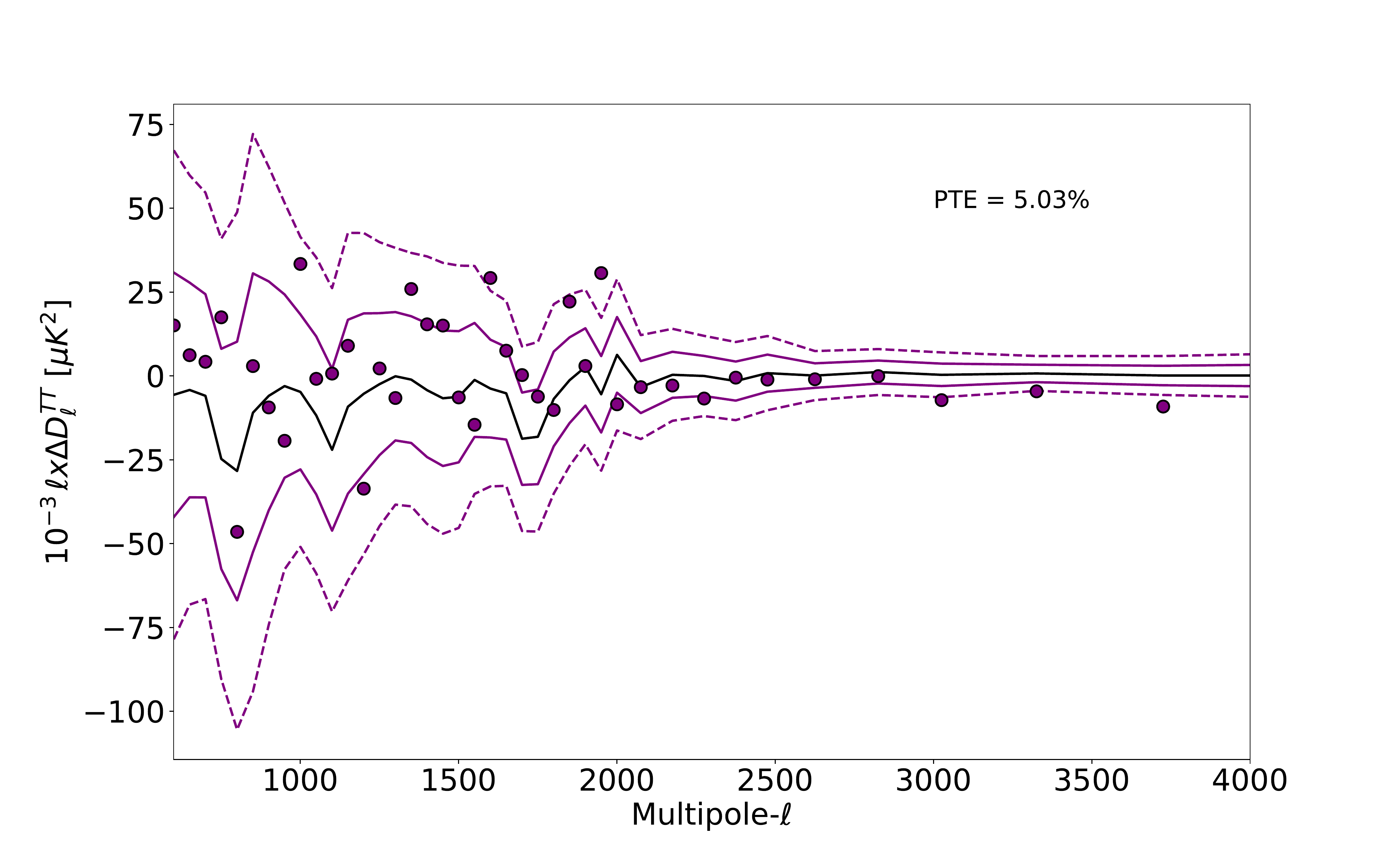}
\vspace{-0.5cm}
\caption{Conditionals for \emph{ACT} $TT$ data given \emph{Planck} $TEEE$ observations. The central line shows the difference with the \emph{Planck} TEEE+low-$\ell$ $\Lambda$CDM best fit. 
}\label{fig:ACT_TT_cond_plc_TEEE}}
\vspace{-0.4cm}
\end{figure}

We show here the most interesting combinations of conditionals that we can run on \emph{Planck}, \emph{ACT} and \emph{SPT} data. More specifically, we compute conditionals for
\begin{itemize}
\item \emph{(a.) P given T within the same experiment:} \emph{Planck} polarization given the \emph{Planck} temperature observations (Fig.~\ref{fig:plc_pol_cond_plc_temp}, with Fig.~\ref{fig:plc_EE_cond_plc_TT} for $EE$ and Fig.~\ref{fig:plc_TE_cond_plc_TT} for $TE$)\footnote{We repeat here these tests for \emph{Planck} for completeness and because as explained later we use slightly different likelihoods than those adopted by the \emph{Planck} team for the conditionals.}; \emph{ACT} polarization given the \emph{ACT} temperature observations (Fig.~\ref{fig:ACT_pol_cond_ACT_temp}, with Fig.~\ref{fig:ACT_EE_cond_ACT_TT} for $EE$ and Fig.~\ref{fig:ACT_TE_cond_ACT_TT} for $TE$).
\item \emph{(b.) T given P within the same experiment:} \emph{Planck} temperature given the \emph{Planck} polarization observations (Fig.~\ref{fig:plc_TT_cond_plc_TEEE}); \emph{ACT} temperature given the \emph{ACT} polarization observations (Fig.~\ref{fig:ACT_TT_cond_ACT_TEEE}).
\item \emph{(c.) P/T given T/P from another experiment:} In this case we adopt \emph{Planck} as benchmark since it is the experiment with the largest constraining power and compute \emph{ACT} polarization given \emph{Planck} temperature observations (Fig.~\ref{fig:ACT_pol_cond_plc_temp}, with Fig.~\ref{fig:ACT_EE_cond_plc_TT} for $EE$ and Fig.~\ref{fig:ACT_TE_cond_plc_TT} for $TE$) and \emph{ACT} temperature given \emph{Planck} polarization (Fig.~\ref{fig:ACT_TT_cond_plc_TEEE}).
\end{itemize}
Finally, we also consider a less stringent test for the \emph{SPT} data. Since \emph{SPT} has only released polarization observations we condition the only two available spectra
\begin{itemize}
    \item \emph{SPT} $TE$ given \emph{SPT} $EE$ and viceversa. 
\end{itemize}


\begin{figure*}[htpb!]
\subfigure[\label{fig:SPT3gEE_condTE}]{\includegraphics[width=0.5\textwidth]{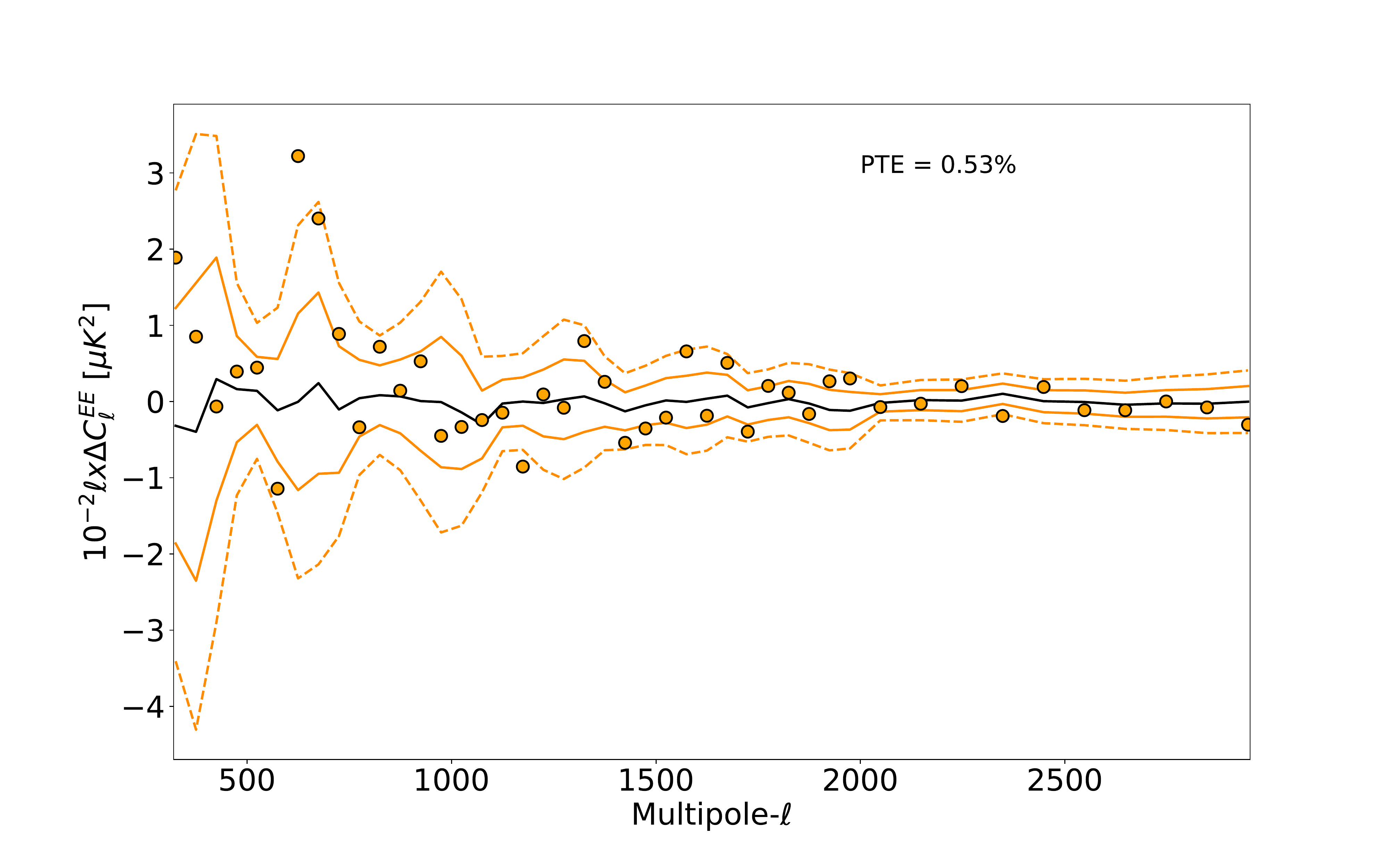}}\hfill
\subfigure[\label{fig:SPT3gTE_condEE}]{\includegraphics[width=0.5\textwidth]{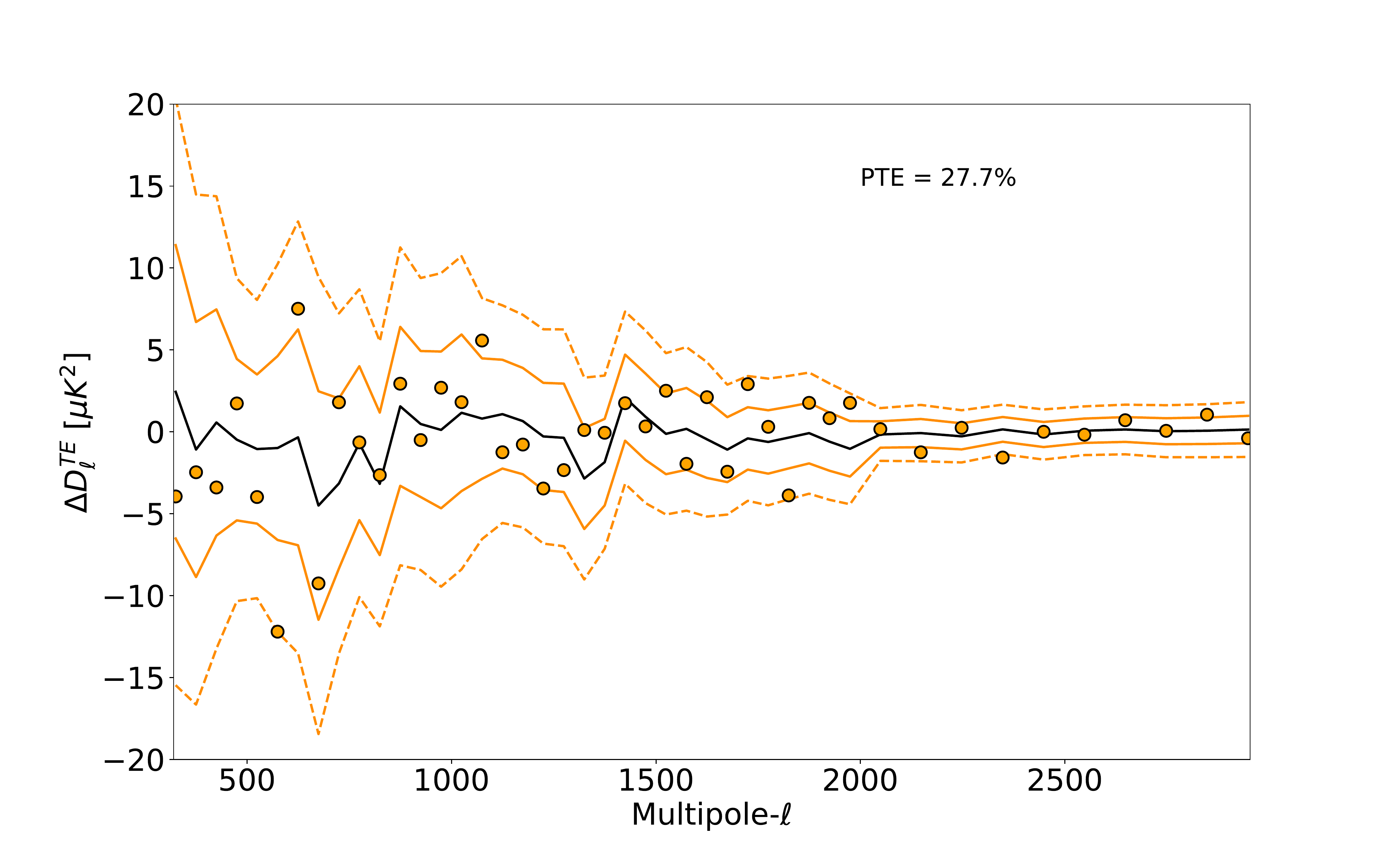}}
\vspace{-0.4cm}
\caption{Conditionals for \emph{SPT} $EE$ \hyperref[fig:SPT3gEE_condTE]{(a)} and $TE$ \hyperref[fig:SPT3gTE_condEE]{(b)} data given \emph{SPT} $TE$ and \emph{SPT} $EE$ observations, respectively. The black line represents the difference between the expected power spectrum given the corresponding observation and the base $\Lambda$CDM best fit determined from \emph{SPT} likelihoods restricted to the polarization counterpart.
}\label{fig:SPT3g_pol_cond_SPT3g_pol}
\vspace{-0.2cm}
\end{figure*}


In what follows we will present and describe the conditionals with a dual purpose. 
We can use this statistic to directly assess and quantify consistency. The main tool used here to perform these tests is the $\chi^2$ in harmonic space, $\chi^2_\mathrm{h}$, defined as 
\begin{equation}\label{eqn:chi2harmonic}
    \chi^2_\mathrm{h} = \sum_{\ell,\ell'=\ell_\mathrm{min}}^{\ell_\mathrm{max}}\left(C_\ell-C_\ell^\mathrm{cond}\right)\mathbf{M}^{-1}_{\ell\ell'}\left(C_{\ell'}-C_{\ell'}^\mathrm{cond}\right)\,,
\end{equation}
where $\mathbf{M}_{\ell\ell'}=\langle\left(C_\ell-C_\ell^\mathrm{cond}\right)\left(C_{\ell'}-C_{\ell'}^\mathrm{cond}\right)\rangle$, $C^\mathrm{cond}_\ell$ represents the conditional mean, and the average is taken over 1000 simulations generated using the covariance matrix of the probability distribution defined in Eq.~\eqref{eqn:clprob}. Secondly, we can use conditionals as a mean to spot unexpected features in the theoretical predictions given a specific observation. These can then be explored in more details with the method presented in Sec.~\ref{sec:consist}.

We also note that our computations are slightly different from the \emph{Planck} conditionals in Ref.s~\cite{Planck2015:cosmo,Planck2018:like}. The \emph{Planck} team derived conditionals using foreground best-fit subtracted spectra and errors. This is an approximation that we relax here using the foreground-marginalised \textsc{plik\_lite} likelihood which fully captures the additional uncertainty due to foregrounds and systematics.
 
Fig.~\ref{fig:plc_pol_cond_plc_temp} shows the conditionals for \emph{Planck} polarization obtained using for the probability distribution in Eq.~\eqref{eqn:thory_from_TT} the theoretical temperature and polarization spectra derived minimizing the \emph{Planck} \textsc{plik\_lite} TT+low-$\ell$\footnote{We remind the reader that all datasets also include a $\tau$ prior but we omit it hereafter in the text and in the captions of the figures to avoid repetitions. We also note that the exact choice of the prior can have small effects.} likelihood, and Eq.s \eqref{eqn:pol_cond_mean} and \eqref{eqn:pol_cond_cov} to compute mean and covariance. The black line of Fig.~\ref{fig:plc_pol_cond_plc_temp} represents the difference between the mean of the polarization conditional given the temperature observations and the base $\Lambda$CDM polarization prediction from the TT+low-$\ell$ best fit. The green lines show the expected covariance for this distribution. More precisely, the solid green and dashed-green lines stand for the 1 and 2$\sigma$ regions. Both probabilities for $EE$ (Fig.~\ref{fig:plc_EE_cond_plc_TT}) and $TE$ (Fig.~\ref{fig:plc_TE_cond_plc_TT}) given temperature observation show an agreement with what the \emph{Planck} team presented in Ref.~\cite{Planck2018:like}. The different treatment of $\tau$ and the additional covariance in \textsc{plik\_lite} move some points but do not affect the overall behaviour. With the exception of a few bins in both $EE$ and $TE$ that show a deviation in power greater than 2$\sigma$, we see good agreement and this is confirmed by the $\chi^2_\mathrm{h}$ for which we find a PTE = 52.04\% for $EE$ and a PTE = 78.45\% for $TE$.

Fig.~\ref{fig:ACT_pol_cond_ACT_temp} shows conditionals for \emph{ACT} polarization given the \emph{ACT} temperature observations -- this is the first time that a conditional analysis is applied to \emph{ACT} data. The theoretical temperature and polarization spectra are derived minimizing the \emph{ACT} TT likelihood within $\Lambda$CDM. The differences between the mean polarization spectra conditioned to temperature observations and the $\Lambda$CDM best-fit spectra are shown by the black lines. The bandpawers used to compute the conditionals are obtained by co-adding the \emph{ACT} deep and wide spectra fully accounting for the correlation between the two regions. The corresponding conditional covariances are shown in the figure with the blue lines. We first note the difference in behaviour between the \emph{Planck} and \emph{ACT} conditionals (i.e., the shape and amplitude of the bands), highlighting the complementary in noise properties, and therefore in constraining power of the two experiments, across multipoles. The plots also prove the capability of \emph{ACT} and other ground-based experiments to perform very stringent tests of the cosmological model at small scales. When looking in detail at the \emph{ACT} $EE$ conditioned to $TT$ (Fig.~\ref{fig:ACT_EE_cond_ACT_TT}) we note $\sim2\sigma$ fluctuation in the conditional covariance band for multipoles $\ell<1000$. This could be a statistical fluctuation or a hint for a poor $\Lambda$CDM fit to the data. Indeed, this feature was described in detail in Ref.~\cite{ACTede} where it was shown that an early dark energy model could accommodate better then $\Lambda$CDM the \emph{ACT} $EE$ data in that region. We have shown here that prior to any extended model fitting we can identify with conditionals regions that are not fully following the expected behaviour in a specific model prediction. More quantitatively, we perform a statistical comparison with the simulations and we find that the PTE obtained with Eq.\eqref{eqn:chi2harmonic} and including only multipoles $\ell<1000$ is 0.25\%, which corresponds to a 3.7 $\chi^2$-excess in terms of $\sqrt{2N_\mathrm{dof}}$. It is worth noting that here we used the standard deviation of a $\chi^2$ distribution due to the low number of points included in the statistics, i.e., $N_\mathrm{dof}=13$ bandpowers. The lowering of the PTE is due to the coherence of the $\sim 2\sigma$ feature, i.e., there is a cumulative contribution to the $\chi^2$ from several multipoles. Note that this also incorporates the preference of higher multipoles because we are cutting the conditional a posteriori from simulations obtained using the full multipole range. Despite this localised feature the overall $EE$ behaviour shows statistical consistency with a PTE = 9.3\%. $TE$ conditionals (Fig.~\ref{fig:ACT_TE_cond_ACT_TT}) show good consistency but with a somewhat low scatter, corresponding to a PTE = 85.32\%.

Fig.~\ref{fig:plc_TT_cond_plc_TEEE} shows the conditional for \emph{Planck} temperature given \emph{Planck} polarization observations. The theoretical temperature and polarization spectra are obtained minimizing the \textsc{plik\_lite} TEEE+low-$\ell$ likelihood. This time Eq.s \eqref{eqn:temp_cond_mean} and \eqref{eqn:temp_cond_cov} are used to compute mean and covariance. The color convention is the same as Fig.~\ref{fig:plc_pol_cond_plc_temp}. Similarly to Ref.~\cite{Planck2018:like}
we find really good agreement between the T/P spectra with a PTE = 55.33\%.

Fig.~\ref{fig:ACT_TT_cond_ACT_TEEE} shows the conditional for \emph{ACT} temperature data given \emph{ACT} polarization observations. Although statistically consistent with zero and a PTE = 25.07\%, the predictions and the bands shows that the \emph{ACT} temperature observations compared to polarization have a tail of low power up to $\ell\sim1500$. As done for $EE$ before, we can explore a specific region by considering only multipoles below $1500$ and we find a PTE = 5.0\% which corresponds to a 1.8 $\chi^2$-excess in terms of $\sqrt{2N_\mathrm{dof}}$.

We now move to conditonials between different experiments. Fig.~\ref{fig:ACT_pol_cond_plc_temp} shows the conditionals of \emph{ACT} polarization given \emph{Planck} temperature observations. A PTE = 21.82\% shows overall good consistency. The fluctuations in $EE$ that we saw earlier are now less pronounced with no points outside the 2$\sigma$ band. The shift in the first bins results in a PTE = 2.0\% when we only consider multipole $\ell<1000$, corresponding to a $\chi^2$-excess in terms of $\sqrt{2N_\mathrm{dof}}$ of 2.44. This reduction of scatter is also seen in Ref.~\cite{ACTede} when one considers the combined \emph{ACT}+\emph{Planck} fit. In $TE$ we notice that the \emph{ACT} spectra show a coherent lack of power at the level of $\sim2\sigma$, with 71\% of the bins below the mean. This was already noticed and explored in the \emph{ACT} analyses~\cite{Aiola2020, ACTede}. As in Ref.~\cite{Aiola2020} our results show that this is similar to a constant $TE$-only amplitude factor, we explore this further in Sec.~\ref{sec:consist}. Despite this feature, we observe an overall statistical consistency with a PTE = 6.49\%.

\emph{ACT} $TT$ conditioned to \emph{Planck} polarization is shown in Fig.~\ref{fig:ACT_TT_cond_plc_TEEE}. Here the conditional shows that \emph{Planck} polarization prefers a somewhat lower $TT$ power spectrum with respect to \emph{ACT} observations, with a resulting PTE = 5.03\%. Most of the contribution to this comes from the high-$\ell$ multipoles ($\ell>2500$), scales not directly probed by \emph{Planck} $TT$ measurements\footnote{When we consider only multipoles at $\ell<2500$ in the $\chi^2_\mathrm{h}$ computation, the PTE increase to 80\%.}. This effect might be reduced when using both \emph{ACT} and \emph{Planck} covariances.

Finally, we look at the \emph{SPT} conditionals in Fig.~\ref{fig:SPT3g_pol_cond_SPT3g_pol} with Fig.~\ref{fig:SPT3gEE_condTE} for $EE$ given $TE$, and Fig.~\ref{fig:SPT3gTE_condEE} for $TE$ given $EE$. For these we use spectra obtained co-adding the three frequency channels, 95, 150 and 220 GHz after subtracting the best-fit foreground and nuisance model\footnote{When using $TE$ to compute $EE$ conditionals we fix the $TE$ nuisance parameters to their best-fit values and the $EE$ parameters to the mean values of the priors -- and viceversa when conditioning $TE$ to $EE$. We checked that using for these parameters the best-fit values from a full TE+EE fit gives very similar results.}. For $EE$ we find that two points are outside the $2\sigma$ region and recover a similar behaviour to the one shown in Fig.~10 of Ref.~\citep{dutcher2021measurements}. The corresponding PTE is 0.52\%, increasing to 2.84\% if we consider only multipoles $\ell>750$. For $TE$ we find a PTE = 27.7\% with no outliers.




\section{Modelling residual transfer functions in CMB polarization}\label{sec:consist}

As long as a model of their effects is available, instrumental systematics (such as polarization efficiency, $T$-to-$E$ leakage, etc.) can be handled during data processing or in the likelihood functions used to put constraints on cosmology. However, even after modelling and marginalizing over these effects, some unresolved and/or unknown systematics can still be present and have an impact. In the previous section we have shown how to use conditional probabilities to localize possible disagreements between different observations. In this section we expand this further and look at ways to then complete the work by modelling and marginalizing over a possible $T-E$ inconsistency. Implementing a model for the $T-E$ inconsistency with this method can help characterize the type and nature of the inconsistency preferred by the data. We will focus on four different models, either inspired by physically motivated systematic effects or by assuming more arbitrary parametrizations. Since these are potentially multipole-dependent deviations from our data model, we consider them as an additional `transfer function', introducing an inconsistency between the CMB temperature and polarization measurements. We further overcome the fact that the conditional analysis performed in Sec.~\ref{sec:cond} does not propagate uncertainties from the best-fit power spectra/cosmology to the conditional distributions. Here, we use a set of extra parameters to constrain the multipole dependence of the deviations and we obtain a joint posterior distribution for them and $\Lambda$CDM parameters. This method should be able to catch either deviations from the $\Lambda$CDM model or unmodelled systematic effects. 

\subsection{Transfer function models}\label{subsec:model}
We introduce four different parametrizations -- $a$ to $d$ below -- to quantify the consistency of temperature and polarization measurements assuming the $\Lambda$CDM model. 
The first two models explore a polarization transfer function and $T$-to-$E$ leakage and are inspired by common CMB systematic effects. The other two models look at possible effects that would alter the $EE$ or $TE$ power spectra separately. In all cases we leave the temperature measurements unaffected by these transfer functions, such that the model for the $TT$ power spectrum corresponds to the actual $TT$ theory power spectrum: $\Tilde{C}_\ell^{TT} = C_\ell^{TT}$. In this way the $TT$ data will help to break the degeneracies between the cosmological parameters and the extra transfer function parameters.\\

\paragraph{Polarization transfer function\\}
We include a transfer function in the polarization $E$ modes ($F_\ell$) such that our model for the observed temperature and polarization modes are $\Tilde{a}_{\ell m}^{T} = a_{\ell m}^{T}$ and $\Tilde{a}_{\ell m}^{E} = F_\ell a_{\ell m}^{E}$ respectively. The model for the measured $TE$ and $EE$ power spectra can be expressed as
\begin{align}\label{eq:modelPE}
    \Tilde{C}_\ell^{TE} &= F_\ell C_\ell^{TE},\nonumber\\
    \Tilde{C}_\ell^{EE} &= F_\ell^2 C_\ell^{EE}.
\end{align}
\vspace{0.01cm}

\paragraph{$T$-to-$E$ leakage\\}
In the case of power leakage between $T$ and $E$ a transfer function, $\beta_\ell$, alters the modes such that $\Tilde{a}_{\ell m}^{T} = a_{\ell m}^T$ and $\Tilde{a}_{\ell m}^E = a_{\ell m}^E + \beta_\ell a_{\ell m}^T$. The measured power spectra are then affected as follows
\begin{align}\label{eq:modelLE}
    \Tilde{C}_\ell^{TE} &= C_\ell^{TE} + \beta_\ell C_\ell^{TT},\nonumber\\
    \Tilde{C}_\ell^{EE} &= C_\ell^{EE} + 2\beta_\ell C_\ell^{TE} + \beta_\ell^2 C_\ell^{TT}.
\end{align}
\vspace{0.01cm}

\paragraph{Independent $EE$ bias\\}
In the case of a transfer function affecting only the $EE$ power spectrum, $\alpha_\ell$, the spectra become
\begin{align}\label{eq:modelEE}
    \Tilde{C}_\ell^{TE} &= C_\ell^{TE},\nonumber\\
    \Tilde{C}_\ell^{EE} &=  \alpha_\ell C_\ell^{EE}.
\end{align}
\vspace{0.01cm}

\paragraph{Independent $TE$ bias\\}
Finally, a transfer function affecting only the $TE$ power spectrum, $\delta_\ell$, leads to
\begin{align}\label{eq:modelTE}
    \Tilde{C}_\ell^{TE} &= \delta_\ell C_\ell^{TE},\nonumber\\
    \Tilde{C}_\ell^{EE} &= C_\ell^{EE}.
\end{align}

With these parametrizations we have four functions to constrain: $F_\ell$, $\beta_\ell$, $\alpha_\ell$ and $\delta_\ell$ -- always considered separately. To study  in a model independent way the multipole dependence of potential deviations from theory that they encode, we use step functions for them with a given number of extra parameters related to the number of available bandpowers in different experiments. We use $n_b=10$ extra parameters for \emph{ACT} and $n_b=11$ for \emph{SPT}; with $N_\mathrm{bins}=40$ for \emph{ACT} power spectra and $N_\mathrm{bins}=44$ for the \emph{SPT} ones, each extra parameter in our model acts on four consecutive CMB bandpowers in the same way. Apart from spotting residual power, this methodology will also capture $T$-$E$ inconsistencies that are localized in a specific multipole range.

\subsection{Results}\label{subsec:results}

We first constrain the self consistency between temperature and polarization measurements within \emph{Planck}, \emph{ACT} and \emph{SPT} latest data. In order to assess the consistency between the \emph{Planck} temperature measurements and the polarization measurements from ground-based experiments, we then also consider some combinations between the $TT$ power spectrum from \emph{Planck} and the $TE$+$EE$ power spectra from \emph{ACT} or \emph{SPT}. Therefore, below we work with the following data combinations 
\begin{itemize}
    \item \emph{Planck} $TT$+$TE$+$EE$,
    \item \emph{ACT} $TT$+$TE$+$EE$, 
    \item \emph{SPT} $TE$+$EE$,
    \item \emph{Planck} $TT$ + \emph{ACT} $TE$+$EE$,
    \item \emph{Planck} $TT$ + \emph{SPT} $TE$+$EE$.
\end{itemize}

Since there is no available measurement of \emph{SPT} temperature, we are limited on the number of tests that we can perform with this dataset. In particular, we cannot study here the case of a polarization transfer function (\emph{a}) and $T-E$ leakage (\emph{b}) for \emph{SPT} data alone.

\begin{figure*}

\subfigure{\includegraphics[width=\textwidth]{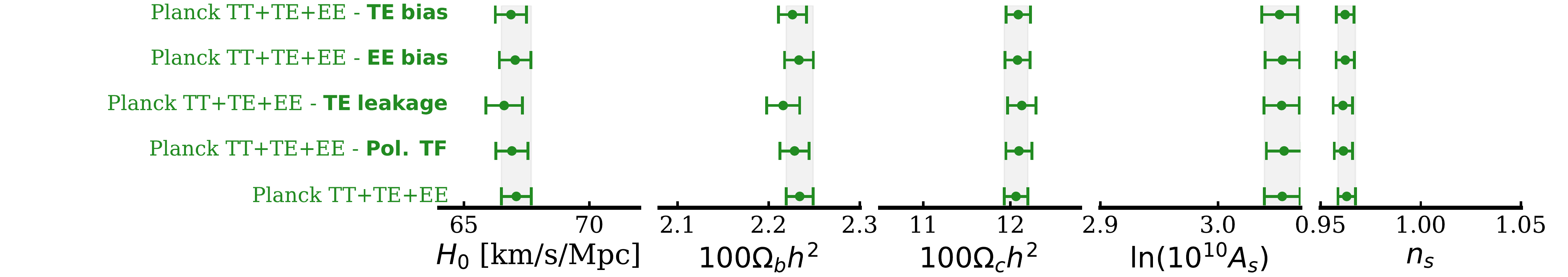}}
\hspace{0mm}
\subfigure{\includegraphics[width=\textwidth]{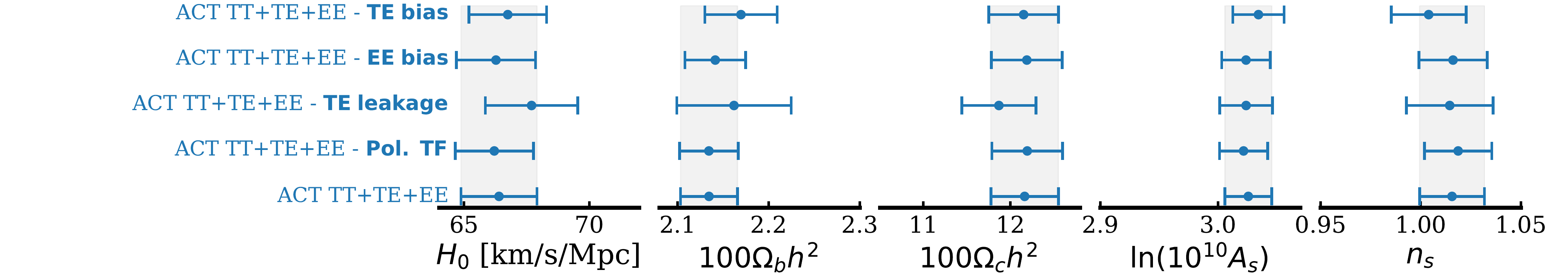}}
\hspace{0mm}
\subfigure{\includegraphics[width=\textwidth]{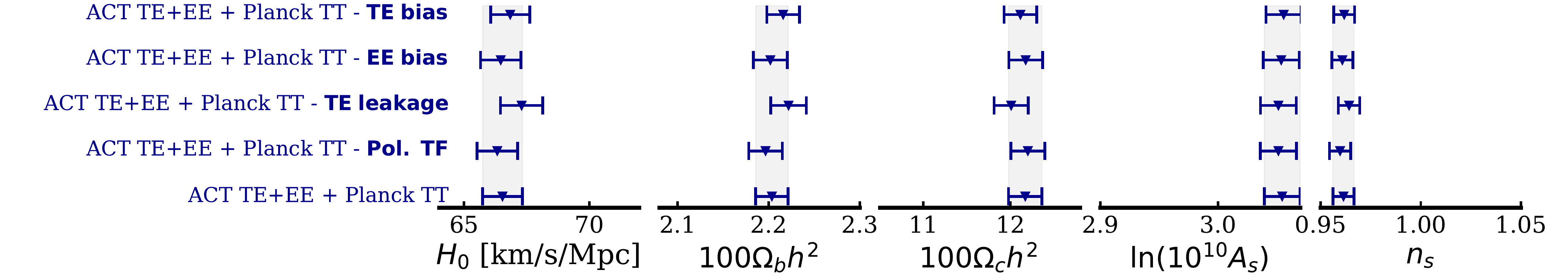}}
\hspace{0mm}
\subfigure{\includegraphics[width=\textwidth]{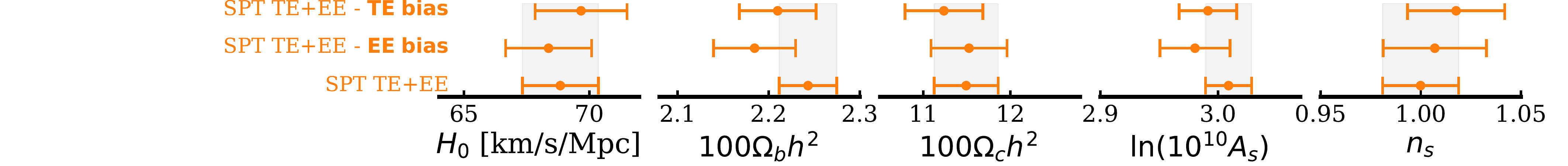}}
\hspace{0mm}
\subfigure{\includegraphics[width=\textwidth]{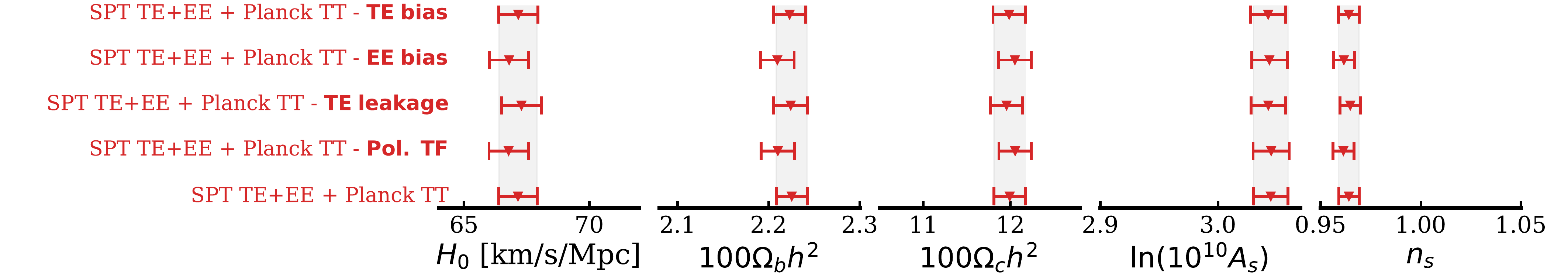}}

\caption{Marginalized constraints on $\Lambda$CDM parameters derived from the different datasets described in Sec.~\ref{subsec:results}: \emph{Planck} $TT$+$TE$+$EE$ (green), \emph{ACT} $TT$+$TE$+$EE$ (lightblue), \emph{SPT} $TE$+$EE$ (orange), \emph{ACT} $TE$+$EE$ + \emph{Planck} $TT$ (blue) and \emph{SPT} $TE$+$EE$ + \emph{Planck} $TT$ (red). We display the $\pm 1\sigma$ (68\%C.L.) constraints on $\Lambda$CDM parameters obtained while also fitting for the transfer function models detailed in Sec.~\ref{subsec:model}. Standard $\Lambda$CDM constraints -- without any additional transfer function -- are displayed at the bottom of each panel and with a vertical grey band.}\label{fig:lcdm_results}
\end{figure*}

As mentioned above, in order to constrain the $n_b$ extra parameters modelling the $T-E$ inconsistencies, we explore the joint posterior distributions of the $n_b$ extra-parameters, the 6 $\Lambda$CDM parameters and the foreground and nuisance parameters described in Sec.~\ref{sec:data} depending on the specific dataset. Parameters are sampled using the MCMC algorithm implemented in \textsc{Cobaya} and marginalized constraints are obtained using \textsc{GetDist}~\cite{getdist}.
Apart from $\tau$, we use flat priors on cosmological and $n_b$ extra parameters. For \emph{SPT} foreground parameters we use the priors described in Ref.~\cite{dutcher2021measurements}. 
Since we are explicitly modelling and fitting for functions that could capture the effect of some systematics that are already accounted for in the likelihoods, to avoid large degeneracy between parameters in some cases we need to change the treatment of the nuisance parameters of the likelihoods. More specifically, we have to fix the polarization efficiencies/calibrations for both \emph{ACT} and \emph{SPT} likelihoods in order to remove some degeneracies between these and the $n_b$ parameters. We use the \textsc{Bobyqa} likelihood maximizer~\cite{Cartis2018a, Cartis2018b} implemented in \textsc{Cobaya} to obtain (within $\Lambda$CDM) the best-fit values: $y_p=1.00047$ for the \emph{ACT} overall polarization efficiency and $E_\mathrm{cal}^{90\mathrm{GHz}}=0.99517$, $E_\mathrm{cal}^{150\mathrm{GHz}}=0.99519$ and $E_\mathrm{cal}^{220\mathrm{GHz}}=1.00073$ for the \emph{SPT} polarization calibrations. We then fix these parameters to their best fits.

We display the 1$\sigma$ constraints on the $\Lambda$CDM parameters derived from \emph{Planck}, \emph{ACT} and \emph{SPT} latest results while fitting at the same time for each of the transfer function models in Fig.~\ref{fig:lcdm_results}. 
We do not see any significant cosmological parameter deviation with respect to the standard analysis (without additional transfer function) also shown in the figure as reference. The $\Lambda$CDM parameters determination is not strongly dependent on the extra parameters describing the $T - E$ inconsistencies (apart from when we lack temperature data as shown later for \emph{SPT}). For all studied cases we observe a preference for high values of the scalar index measured from ground-based experiments, and we recover a preference for lower values of $n_s$ when combining with \emph{Planck} temperature data.

\begin{table*}[hbp!]
\begin{ruledtabular}
\begin{tabular}{lcccc}
 &\multicolumn{4}{c}{$\chi^2$/d.o.f (PTE)}\\
Dataset & Pol. TF & $T$-to-$E$ leakage & $EE$ bias & $TE$ bias \\
\hline
\emph{Planck} TT+TE+EE & 6.74/10 (0.75) & 14.2/10 (0.16) & 8.07/10 (0.62) & 12.6/10 (0.25)\\
\emph{ACT} TT+TE+EE & 8.91/10 (0.54) & 4.76/10 (0.91) & 10.52/10 (0.40) & 6.61/10 (0.76)\\
\emph{Planck} TT + \emph{ACT} TE+EE & 11.29/10 (0.34) & 15.30/10 (0.12) & 11.35/10 (0.33) & 17.82/10 (0.06)\\
\emph{SPT} TE+EE & $$---$$ & $$---$$ & 14.38/11 (0.21) & 11.69/11 (0.39)\\
\emph{Planck} TT + \emph{SPT} TE+EE & 18.26/11 (0.08) & 8.89/11 (0.63) & 16.53/11 (0.12) & 10.32/11 (0.50)
\end{tabular}
\end{ruledtabular}
\caption{\label{tab:chi2} Goodness of the fit, in terms of $\chi^2$ and probability to exceed (PTE), for all for the different datasets and transfer function models considered in this analysis.}
\end{table*}

\begin{figure*}[htp!]
\subfigure[~Polarization transfer function\label{fig:poleff}]{\includegraphics[width=0.5\textwidth]{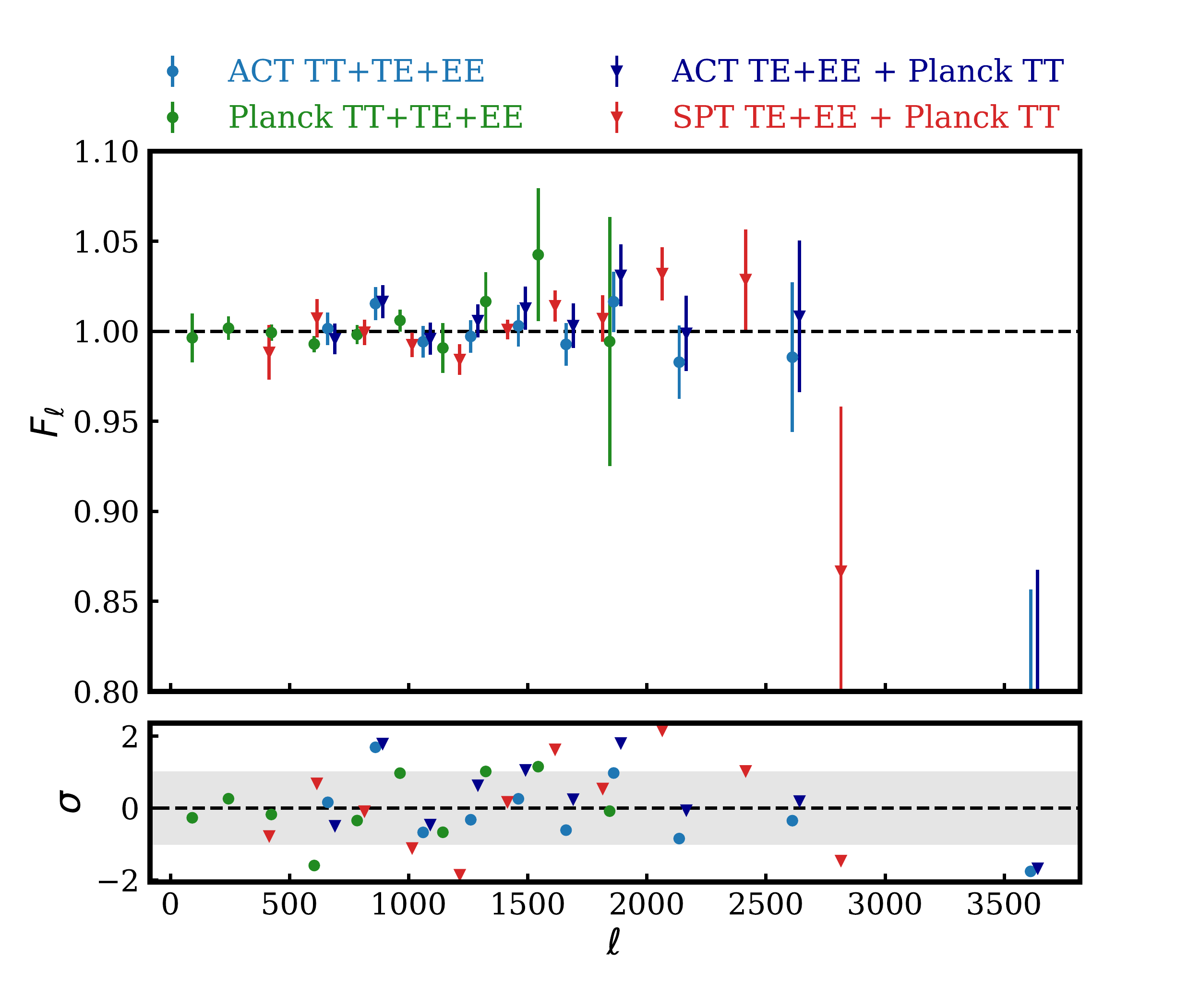}}\hfill
\subfigure[~$T$ to $E$ leakage\label{fig:leak}]{\includegraphics[width=0.5\textwidth]{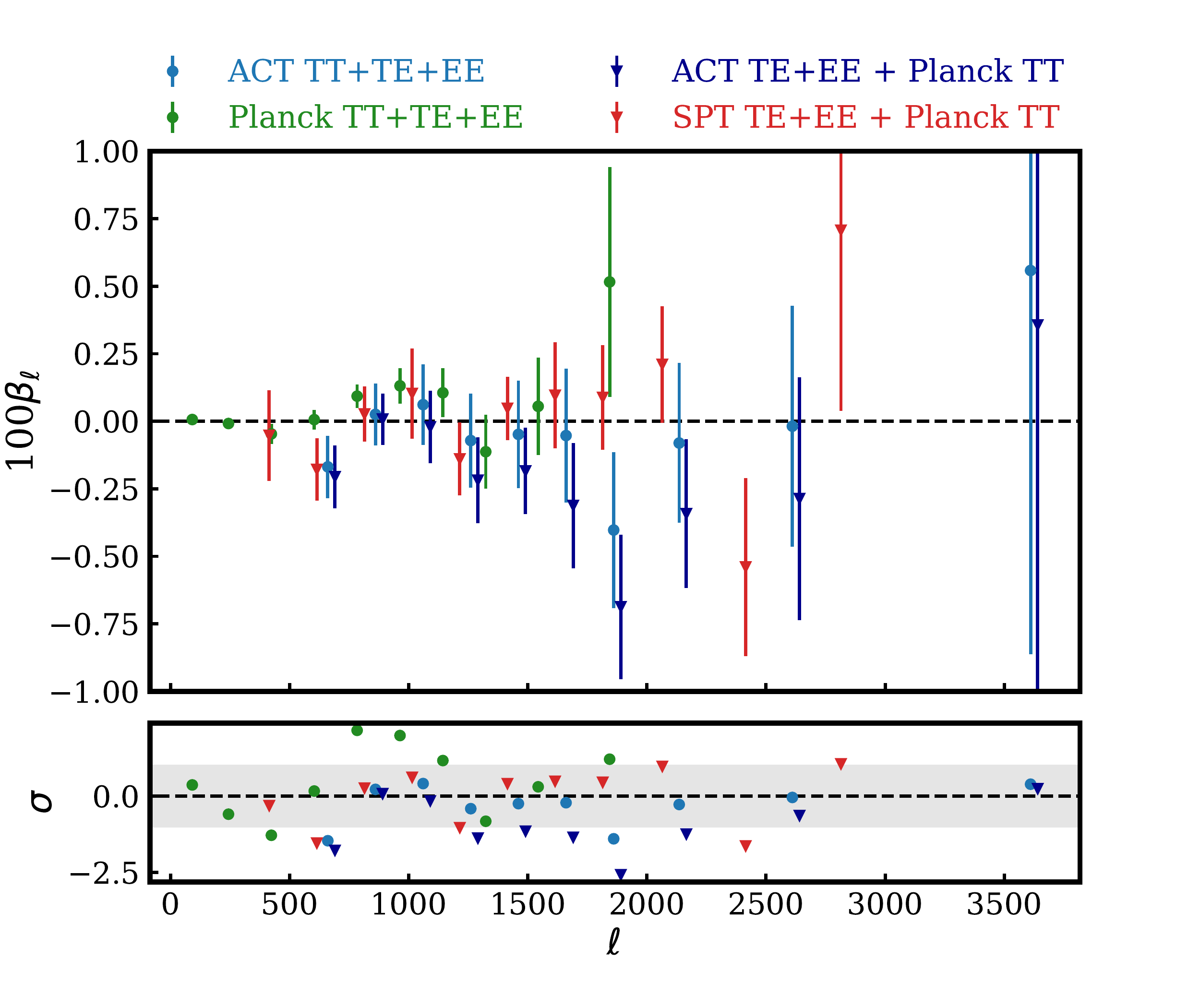}}
\hspace{0mm}
\subfigure[~$EE$ bias\label{fig:eecrap}]{\includegraphics[width=0.5\textwidth]{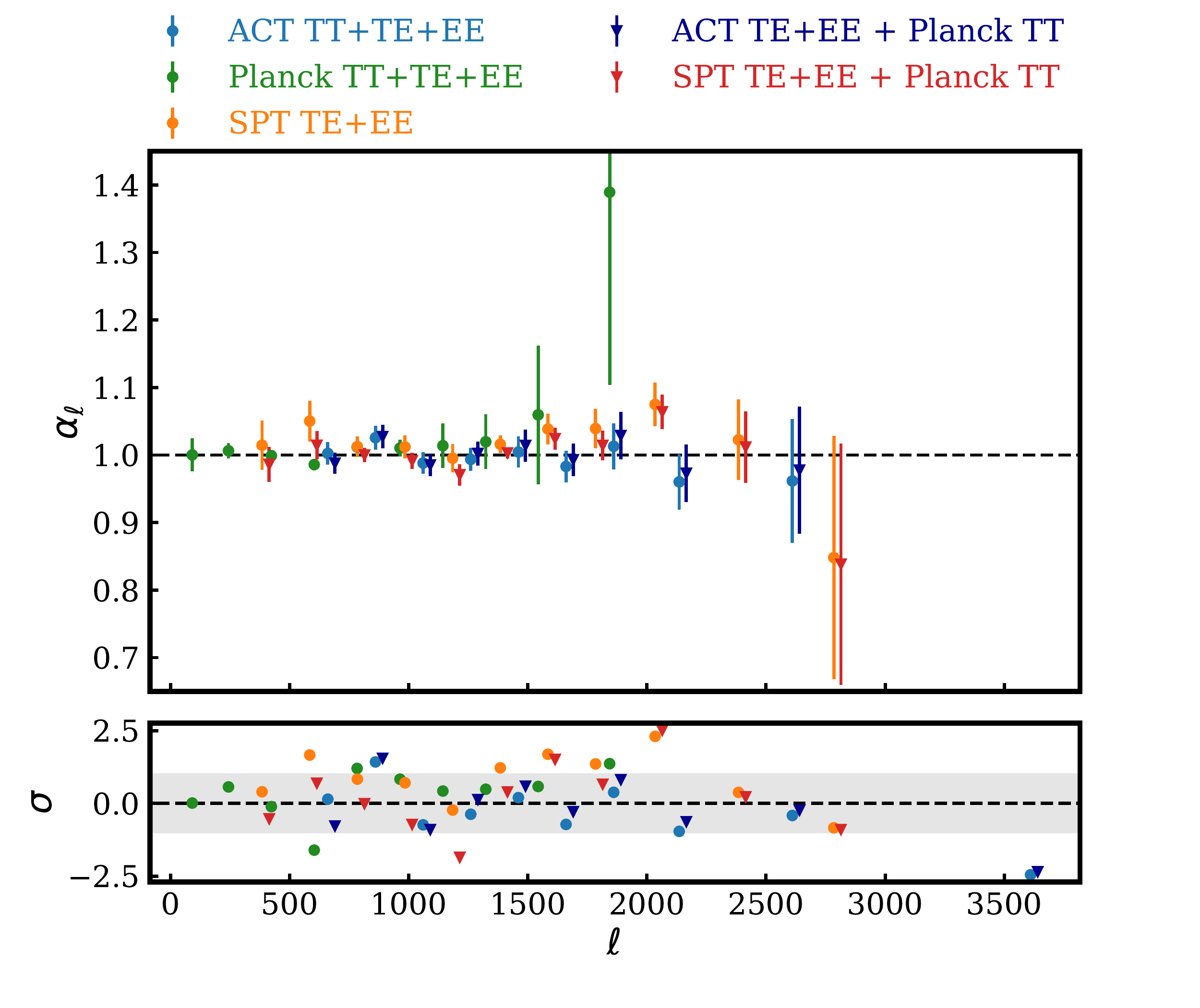}}\hfill
\subfigure[~$TE$ bias\label{fig:tecrap}]{\includegraphics[width=0.5\textwidth]{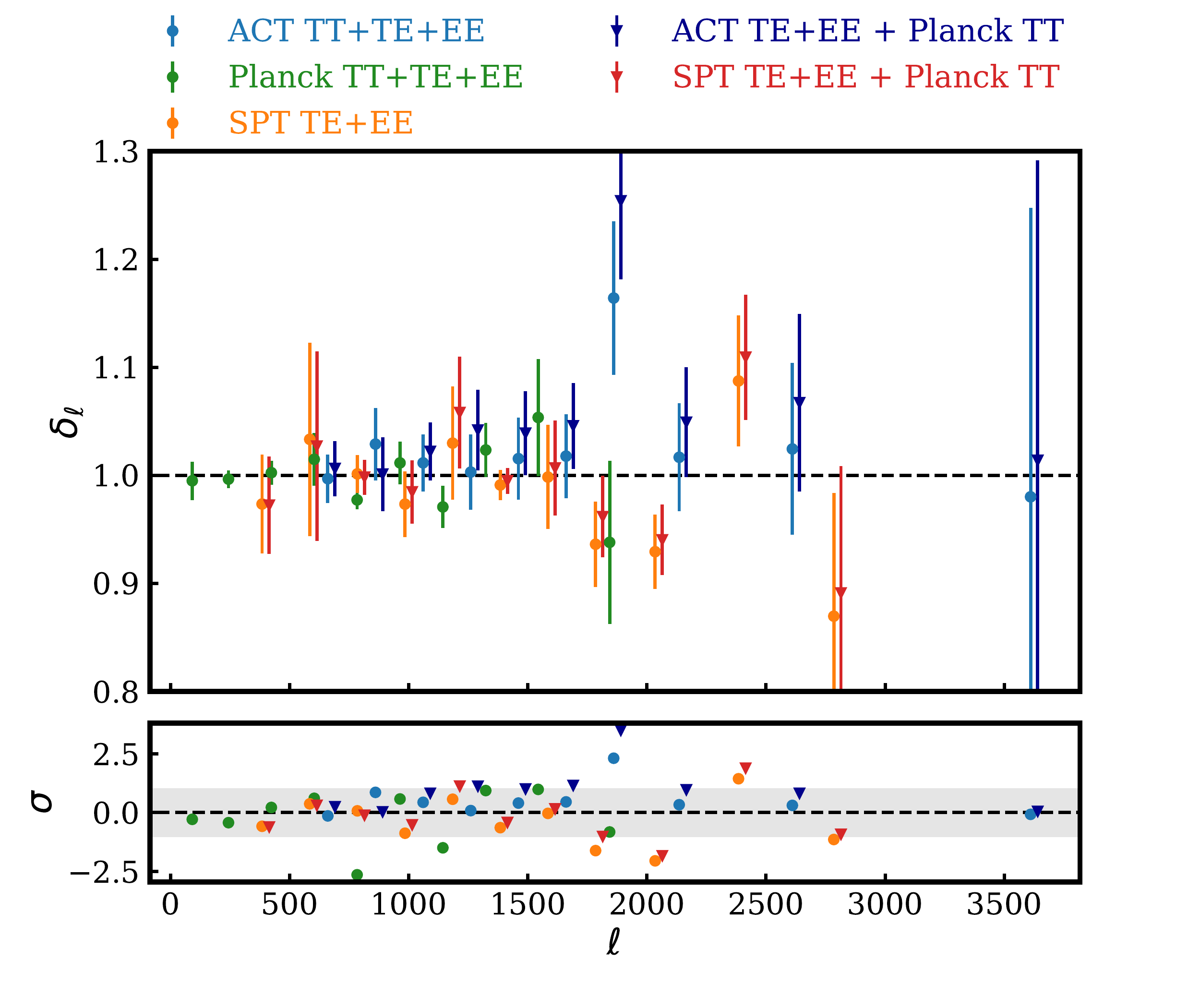}}
\vspace{-0.3cm}
\caption{Transfer function bandpowers with $\pm$1$\sigma$ errors for the four different models studied: the $F_\ell$ polarization transfer function \hyperref[fig:poleff]{(a)}, the $\beta_\ell$ $T$-to-$E$ leakage \hyperref[fig:leak]{(b)}, the $\alpha_\ell$ $EE$ bias \hyperref[fig:eecrap]{(c)} and the $\delta_\ell$ $TE$ bias \hyperref[fig:tecrap]{(d)}. On each panel we display the results for different datasets: Planck 2018 $TT$+$TE$+$EE$ (green), \emph{ACT} $TT$+$TE$+$EE$ (lightblue), \emph{SPT} $TE$+$EE$ (orange), \emph{Planck} $TT$ + \emph{ACT} $TE$+$EE$ (blue) and \emph{Planck} $TT$ + \emph{SPT} $TE$+$EE$ (red). The lower panels of each sub-figure show the comparison with, or potential deviation from, the expected value in units of $\sigma$. The gray band corresponds to the $\pm$1$\sigma$ limits. No statistically-significant deviation is observed.}
\vspace{-0.1cm}
\end{figure*}

\begin{figure*}[t!]
\subfigure[~\emph{SPT} $TE$+$EE$\label{fig:spt_corrA}]{\includegraphics[width=0.5\textwidth]{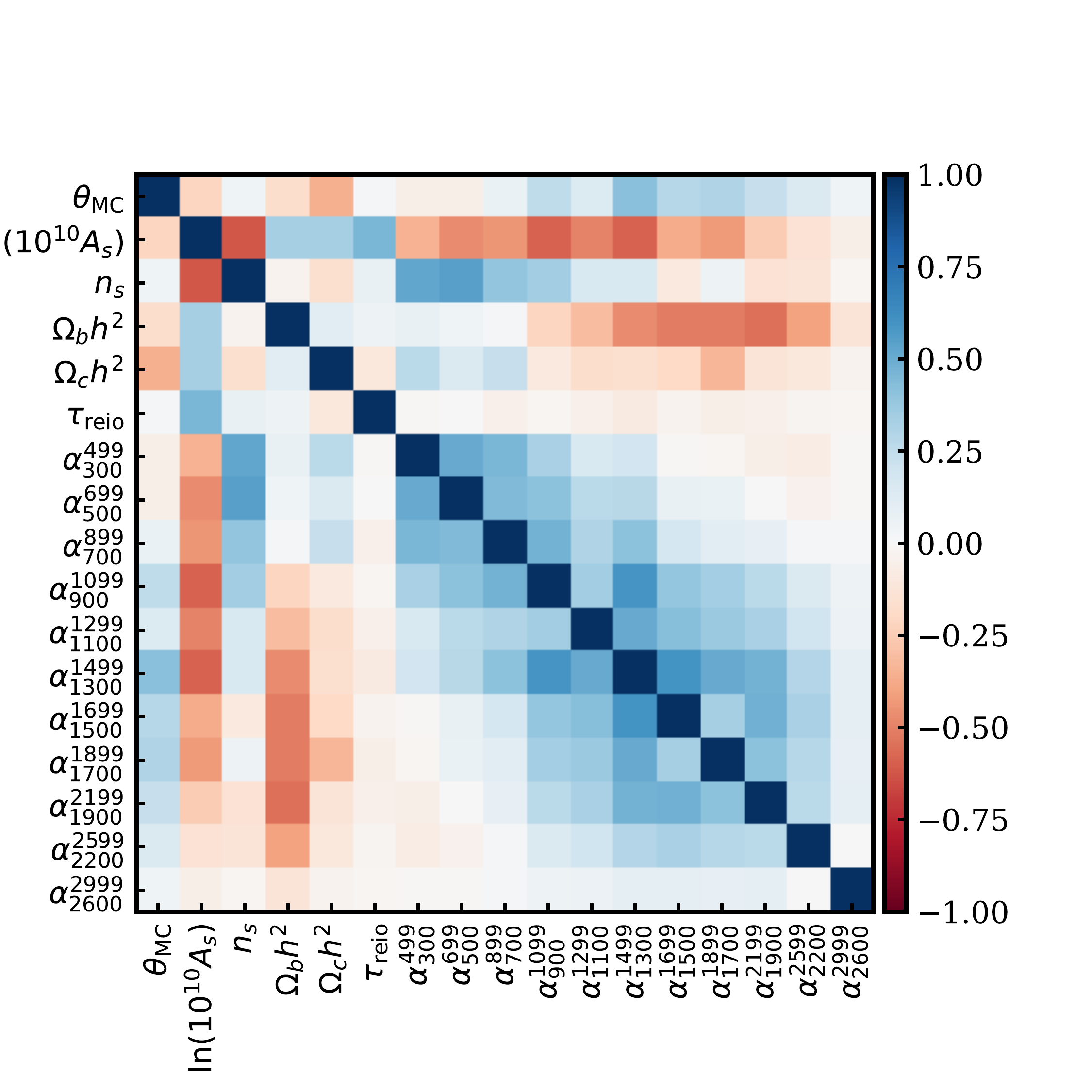}}\hfill
\subfigure[~\emph{SPT} $TE$+$EE$ + \emph{Planck} $TT$\label{fig:spt_corrB}]{\includegraphics[width=0.5\textwidth]{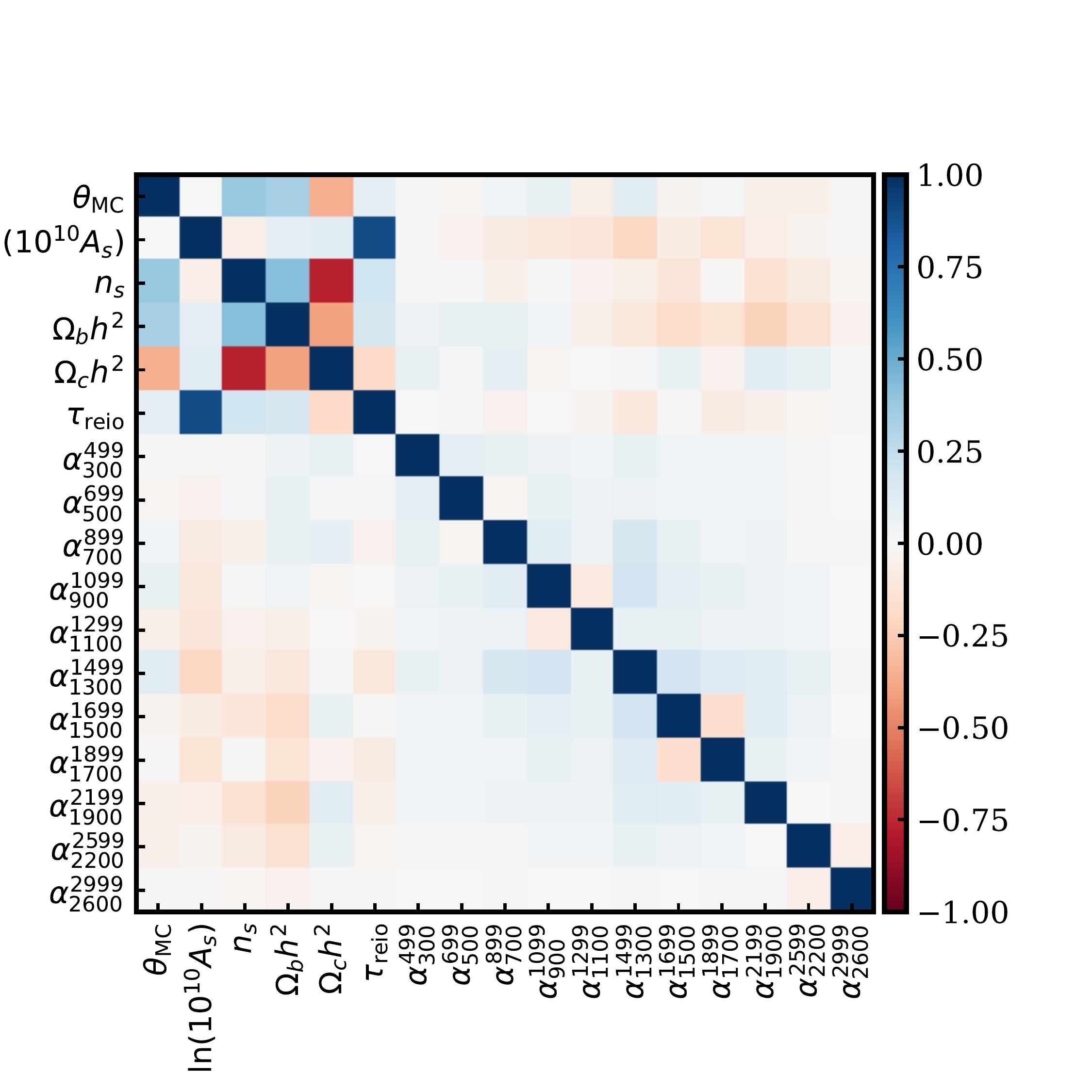}}
\caption{Correlation matrices of cosmological parameters and transfer function bandpowers obtained using a $EE$ transfer function. We display on the left the correlation matrix derived from \emph{SPT} ($TE$+$EE$) data [\hyperref[fig:spt_corrA]{(a)}] and on the right the correlation matrix derived from a combination of \emph{SPT} ($TE$+$EE$) and \emph{Planck} 2018 $TT$ data [\hyperref[fig:spt_corrB]{(b)}].}
\end{figure*}\label{fig:spt_corr}

In order to quantify the deviation with respect to theory expectations, we compare the $n_b$ constraints to the expected value for the studied bias (i.e., 0 for $T$-to-$E$ leakage and 1 otherwise). We then compute a $\chi^2$ for each dataset and bias model using the parameter covariance matrix obtained from the MCMC chains. The results are displayed in Table~\ref{tab:chi2}. Figures~\ref{fig:poleff}, \ref{fig:leak}, \ref{fig:eecrap} and~\ref{fig:tecrap} display the 1$\sigma$ constraints on the transfer function bandpowers and compares them with the expectation in the case of consistency.
We find no evidence for a polarization transfer function in both \emph{Planck} and \emph{ACT} data (Fig.~\ref{fig:poleff}), with PTE=75\%, 54\% respectively. We observe a small feature around $\ell\simeq 900$ in the $T$-to-$E$ leakage transfer function estimated from \emph{Planck} data, but this deviation is not statistically significant with a PTE=16\% (Fig.~\ref{fig:leak}). The $EE$/$TE$ transfer functions derived from \emph{Planck}, \emph{ACT} and \emph{SPT} data are also in good agreement with expectations with PTE=62\%, 40\% and 21\% ($EE$ transfer function) and PTE=25\%, 76\% and 39\% ($TE$ transfer function).

When we quantify the $T-E$ inconsistencies between polarization measurements from ground-based experiments (\emph{ACT}, \emph{SPT}) and temperature measurements from \emph{Planck}, we observe a slight degradation in $\chi^2$ with respect to the values obtained considering \emph{Planck}, \emph{ACT} and \emph{SPT} individually (except for the $TE$ transfer function constrained from \emph{SPT} data) but still no significant deviation from consistency. We note one case in particular. For the combination of \emph{Planck} $TT$ and \emph{ACT} $TE$+$EE$, we recover a slight preference for higher values of the $TE$ power spectrum with respect to what the $\Lambda$CDM model predicts, with a PTE=6\%, and mostly driven by the large value of the transfer function in the bin centered on $\ell=1875.5$. As mentioned above, this effect was noticed in Ref.~\cite{Aiola2020} where an artificial 5\% effect in the $TE$ calibration was explored and discussed, and also seen in Fig.~\ref{fig:ACT_EE_cond_plc_TT} of Sec.~\ref{sec:cond}. We run additional tests here to quantify this feature in more detail. We define and constrain three independent calibration amplitudes $A_{TT}$, $A_{EE}$ and $A_{TE}$ for \emph{ACT} DR4 ($C_\ell^{TT}$, $C_\ell^{EE}$, $C_\ell^{TE}$) and including also the large scale temperature measurements from \emph{Planck} ($C_\ell^{TT}$ at $\ell<650$). While for $TT$ and $EE$ we find no particular preference away from unity, for $TE$ we obtain a marginalized constraint $A_{TE} = 1.037 \pm 0.015$ at 68\% confidence which is 2.5$\sigma$ away from the standard, no-inconsistency, value of 1. Even if visible in our results, we note that this is a small deviation with respect to $\Lambda$CDM. Overall, we find no statistically significant evidence for transfer functions.

Finally we highlight the importance of temperature data for fitting these systematic models. We show the correlation matrices between $\Lambda$CDM and the parameters describing the shape of the $EE$ transfer function for \emph{SPT} $TE$+$EE$ alone (Fig.~\ref{fig:spt_corrA}) or for a combination of \emph{SPT} data with \emph{Planck} $TT$ data (Fig.~\ref{fig:spt_corrB}). In the first case, we do not have any measurement of the temperature power spectrum and we observe non-zero correlations between the extra parameters and cosmological parameters even if the $TE$ power spectrum is unchanged when we constrain the $EE$ transfer function. These correlations are much smaller when we include temperature data from \emph{Planck}.

\section{Conclusion}\label{sec:Conclusion}

In this work we have presented methods to quantify the consistency between CMB temperature and polarization measurements and applied them to the most recent data from \emph{Planck}, \emph{ACT} and \emph{SPT}. 
\begin{itemize}
    \item We have performed a full survey of the datasets with conditional probabilities in Sec.~\ref{sec:cond} which have been compared to simulations and show good agreement between temperature and polarization within the same experiments, as well as between different experiments.
    \item We have studied potential $T-E$ inconsistencies directly modelling and fitting for transfer functions in Sec.~\ref{sec:consist}. We constrained the extra parameters introduced to model the transfer functions together with cosmological parameters. Again, we found no evidence for an inconsistency within \emph{ACT}, \emph{SPT}, and \emph{Planck} or between the ground-based polarization data and the \emph{Planck} temperature measurements.
\end{itemize}

This work introduced a number of methods to potentially spot some deviations from the $\Lambda$CDM predictions either due to physics beyond the standard model or due to instrumental systematic effects. While we found no statistically-significant evidence for such deviations in current CMB data, the accuracy of future small-scale polarization data from \emph{ACT}, \emph{SPT} and the Simons Observatory will allow to apply these methods in a much more stringent way and potentially identify and study inconsistencies between CMB temperature and polarization with high significance.

\subsection*{Acknowledgements}
The theoretical power spectra used in this paper were computed using \texttt{CAMB} Boltzmann solver~\cite{Lewis2000, Howlett2012}. We thank Edward J. Wollack for useful comments. We gratefully acknowledge the IN2P3 Computer Center (\url{http://cc.in2p3.fr}) and the Hawk high-performance computing cluster at the Advanced Research Computing at Cardiff (ARCCA) for providing the computing resources and services needed to this work. EC acknowledges support from the STFC Ernest Rutherford Fellowship ST/M004856/2 and STFC Consolidated Grant ST/S00033X/1.  EC and UN acknowledge support from the European Research Council (ERC) under the European Union’s Horizon 2020 research and innovation programme (Grant agreement No. 849169).\\

\begin{appendix}
\section{SPT-3G likelihood}\label{app:SPT3G}
\begin{figure*}[htp!]
    \includegraphics[width=\textwidth]{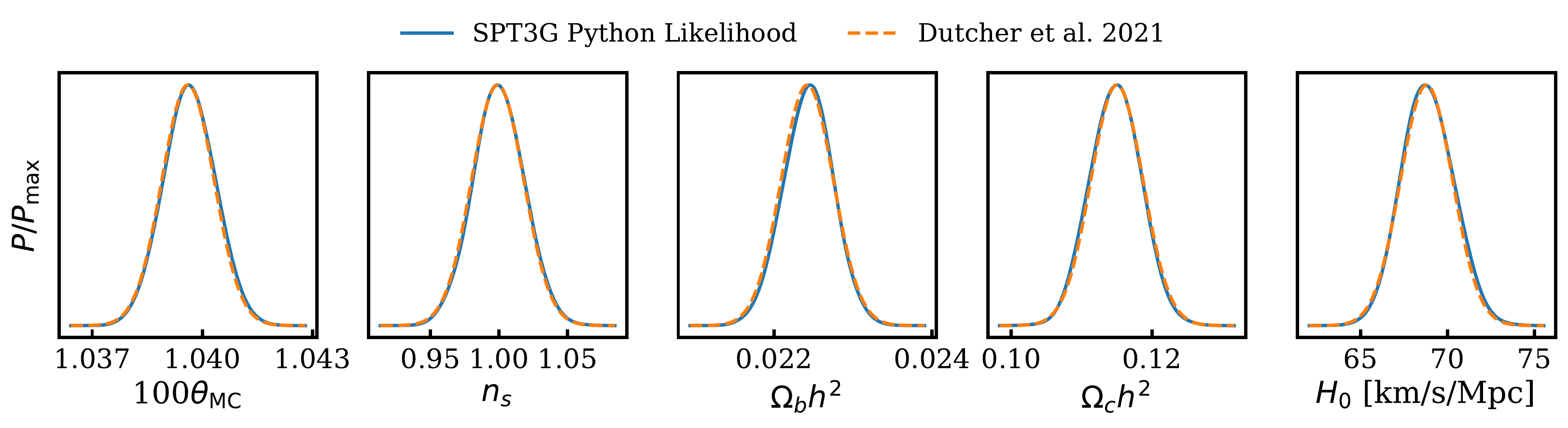}
    \caption{$\Lambda$CDM parameters posterior distributions derived using the \textsc{python} implementation of the SPT-3G likelihood (solid blue) compared to the official SPT-3G collaboration constraints from Ref.~\cite{dutcher2021measurements} (dashed orange).}
    \label{fig:app_post_spt}
\end{figure*}

In this section we validate the results of our SPT-3G python likelihood with respect to the results published in Ref.~\cite{dutcher2021measurements}. Since our SPT-3G likelihood is compatible with \textsc{Cobaya}, we obtain the posterior distributions displayed in Fig.~\ref{fig:app_post_spt} using \textsc{Cobaya}, computing the theory power spectra with \textsc{Camb} (with the default accuracy settings and \texttt{lens\_potential\_accuracy=1.0}). For the MCMC analysis, we set flat priors on cosmological parameters except for the reionization optical depth $\tau$ for which we use $\tau = 0.0544 \pm 0.0073$. We use the Gaussian priors from Ref.~\cite{dutcher2021measurements} for the point sources parameters ($D^{\mathrm{ps}, \nu_1\times\nu_2}$), the parameters describing the polarized galactic dust emissions in $EE$ and $TE$ ($A_\mathrm{d}^{EE}$, $\alpha_\mathrm{d}^{EE}$, $A_\mathrm{d}^{TE}$, $\alpha_\mathrm{d}^{TE}$) and the mean lensing convergence $\kappa$. We impose flat priors on the temperature and polarization map calibration parameters $(T_\mathrm{cal}^\nu, E_\mathrm{cal}^\nu) \in [0.85, 1.15]^2$.
The 68\% constraints obtained with our likelihood are displayed in Table~\ref{tab:cosmo_spt}. We recover well the published constraints on cosmology. We compute a $\chi^2$ using cosmological parameters from Table~\ref{tab:cosmo_spt}, and obtained  $\chi^2 = 513.5$ for $528$ bandpowers ($\chi2 = 513.0/528$ in Ref.~\cite{dutcher2021measurements})
\end{appendix}

\begin{table}[htpb]
\begin{ruledtabular}
\begin{tabular}{lcc}
& SPT-3G (this work) & SPT-3G~\cite{dutcher2021measurements} \\
\hline
100$\theta_\mathrm{MC}$ & $1.03965 \pm 0.00072$ & $1.03961 \pm 0.00071$\\
$\Omega_bh^2$ & $0.02243 \pm 0.00032$ & $0.02242 \pm 0.00033$\\
$\Omega_ch^2$ & $0.1148 \pm 0.0037$ & $0.1150 \pm 0.0037$\\
$n_s$ & $1.000 \pm 0.019$ & $0.999 \pm 0.019$\\
$H_0$ [km/s/Mpc] & $68.9 \pm 1.5$ & $68.8 \pm 1.5$
\end{tabular}
\end{ruledtabular}
\caption{\label{tab:cosmo_spt} Constraints and 68\% errors on $\Lambda$CDM parameters from SPT-3G using our \textsc{python} likelihood compared with the constraints from Ref.~\cite{dutcher2021measurements}}
\end{table}

\bibliography{draft}

\begin{thebibliography}{24}
\expandafter\ifx\csname natexlab\endcsname\relax\def\natexlab#1{#1}\fi
\expandafter\ifx\csname bibnamefont\endcsname\relax
  \def\bibnamefont#1{#1}\fi
\expandafter\ifx\csname bibfnamefont\endcsname\relax
  \def\bibfnamefont#1{#1}\fi
\expandafter\ifx\csname citenamefont\endcsname\relax
  \def\citenamefont#1{#1}\fi
\expandafter\ifx\csname url\endcsname\relax
  \def\url#1{\texttt{#1}}\fi
\expandafter\ifx\csname urlprefix\endcsname\relax\def\urlprefix{URL }\fi
\providecommand{\bibinfo}[2]{#2}
\providecommand{\eprint}[2][]{\url{#2}}

\bibitem[{\citenamefont{{Planck Collaboration VI}}(2020)}]{Planck2018:cosmo}
\bibinfo{author}{\bibnamefont{{Planck Collaboration VI}}},
  \bibinfo{journal}{A\&A} \textbf{\bibinfo{volume}{641}}, \bibinfo{pages}{A6}
  (\bibinfo{year}{2020}),
  \urlprefix\url{https://doi.org/10.1051/0004-6361/201833910}.

\bibitem[{\citenamefont{{Aiola} et~al.}(2020)\citenamefont{{Aiola},
  {Calabrese}, {Maurin}, {Naess}, {Schmitt}, {Abitbol}, {Addison}, {Ade},
  {Alonso}, {Amiri} et~al.}}]{Aiola2020}
\bibinfo{author}{\bibfnamefont{S.}~\bibnamefont{{Aiola}}},
  \bibinfo{author}{\bibfnamefont{E.}~\bibnamefont{{Calabrese}}},
  \bibinfo{author}{\bibfnamefont{L.}~\bibnamefont{{Maurin}}},
  \bibinfo{author}{\bibfnamefont{S.}~\bibnamefont{{Naess}}},
  \bibinfo{author}{\bibfnamefont{B.~L.} \bibnamefont{{Schmitt}}},
  \bibinfo{author}{\bibfnamefont{M.~H.} \bibnamefont{{Abitbol}}},
  \bibinfo{author}{\bibfnamefont{G.~E.} \bibnamefont{{Addison}}},
  \bibinfo{author}{\bibfnamefont{P.~A.~R.} \bibnamefont{{Ade}}},
  \bibinfo{author}{\bibfnamefont{D.}~\bibnamefont{{Alonso}}},
  \bibinfo{author}{\bibfnamefont{M.}~\bibnamefont{{Amiri}}},
  \bibnamefont{et~al.}, \bibinfo{journal}{\jcap}
  \textbf{\bibinfo{volume}{2020}}, \bibinfo{eid}{047} (\bibinfo{year}{2020}),
  \eprint{2007.07288}.

\bibitem[{\citenamefont{Dutcher et~al.}(2021)\citenamefont{Dutcher, Balkenhol,
  Ade, Ahmed, Anderes, Anderson, Archipley, Avva, Aylor, Barry
  et~al.}}]{dutcher2021measurements}
\bibinfo{author}{\bibfnamefont{D.}~\bibnamefont{Dutcher}},
  \bibinfo{author}{\bibfnamefont{L.}~\bibnamefont{Balkenhol}},
  \bibinfo{author}{\bibfnamefont{P.~A.~R.} \bibnamefont{Ade}},
  \bibinfo{author}{\bibfnamefont{Z.}~\bibnamefont{Ahmed}},
  \bibinfo{author}{\bibfnamefont{E.}~\bibnamefont{Anderes}},
  \bibinfo{author}{\bibfnamefont{A.~J.} \bibnamefont{Anderson}},
  \bibinfo{author}{\bibfnamefont{M.}~\bibnamefont{Archipley}},
  \bibinfo{author}{\bibfnamefont{J.~S.} \bibnamefont{Avva}},
  \bibinfo{author}{\bibfnamefont{K.}~\bibnamefont{Aylor}},
  \bibinfo{author}{\bibfnamefont{P.~S.} \bibnamefont{Barry}},
  \bibnamefont{et~al.} (\bibinfo{year}{2021}), \eprint{2101.01684}.

\bibitem[{\citenamefont{{Balkenhol} et~al.}(2021)\citenamefont{{Balkenhol},
  {Dutcher}, {Ade}, {Ahmed}, {Anderes}, {Anderson}, {Archipley}, {Avva},
  {Aylor}, {Barry} et~al.}}]{Balkenhol2021}
\bibinfo{author}{\bibfnamefont{L.}~\bibnamefont{{Balkenhol}}},
  \bibinfo{author}{\bibfnamefont{D.}~\bibnamefont{{Dutcher}}},
  \bibinfo{author}{\bibfnamefont{P.~A.~R.} \bibnamefont{{Ade}}},
  \bibinfo{author}{\bibfnamefont{Z.}~\bibnamefont{{Ahmed}}},
  \bibinfo{author}{\bibfnamefont{E.}~\bibnamefont{{Anderes}}},
  \bibinfo{author}{\bibfnamefont{A.~J.} \bibnamefont{{Anderson}}},
  \bibinfo{author}{\bibfnamefont{M.}~\bibnamefont{{Archipley}}},
  \bibinfo{author}{\bibfnamefont{J.~S.} \bibnamefont{{Avva}}},
  \bibinfo{author}{\bibfnamefont{K.}~\bibnamefont{{Aylor}}},
  \bibinfo{author}{\bibfnamefont{P.~S.} \bibnamefont{{Barry}}},
  \bibnamefont{et~al.}, \bibinfo{journal}{\prd} \textbf{\bibinfo{volume}{104}},
  \bibinfo{eid}{083509} (\bibinfo{year}{2021}), \eprint{2103.13618}.

\bibitem[{\citenamefont{{Planck Collaboration}
  et~al.}(2020)\citenamefont{{Planck Collaboration}, {Aghanim, N.}, {Akrami,
  Y.}, {Ashdown, M.}, {Aumont, J.}, {Baccigalupi, C.}, {Ballardini, M.},
  {Banday, A. J.}, {Barreiro, R. B.}, {Bartolo, N.} et~al.}}]{P18_like}
\bibinfo{author}{\bibnamefont{{Planck Collaboration}}},
  \bibinfo{author}{\bibnamefont{{Aghanim, N.}}},
  \bibinfo{author}{\bibnamefont{{Akrami, Y.}}},
  \bibinfo{author}{\bibnamefont{{Ashdown, M.}}},
  \bibinfo{author}{\bibnamefont{{Aumont, J.}}},
  \bibinfo{author}{\bibnamefont{{Baccigalupi, C.}}},
  \bibinfo{author}{\bibnamefont{{Ballardini, M.}}},
  \bibinfo{author}{\bibnamefont{{Banday, A. J.}}},
  \bibinfo{author}{\bibnamefont{{Barreiro, R. B.}}},
  \bibinfo{author}{\bibnamefont{{Bartolo, N.}}}, \bibnamefont{et~al.},
  \bibinfo{journal}{A\&A} \textbf{\bibinfo{volume}{641}}, \bibinfo{pages}{A5}
  (\bibinfo{year}{2020}),
  \urlprefix\url{https://doi.org/10.1051/0004-6361/201936386}.

\bibitem[{\citenamefont{Choi et~al.}(2020)\citenamefont{Choi, Hasselfield, Ho,
  Koopman, Lungu, Abitbol, Addison, Ade, Aiola, Alonso et~al.}}]{Choi2020}
\bibinfo{author}{\bibfnamefont{S.~K.} \bibnamefont{Choi}},
  \bibinfo{author}{\bibfnamefont{M.}~\bibnamefont{Hasselfield}},
  \bibinfo{author}{\bibfnamefont{S.-P.~P.} \bibnamefont{Ho}},
  \bibinfo{author}{\bibfnamefont{B.}~\bibnamefont{Koopman}},
  \bibinfo{author}{\bibfnamefont{M.}~\bibnamefont{Lungu}},
  \bibinfo{author}{\bibfnamefont{M.~H.} \bibnamefont{Abitbol}},
  \bibinfo{author}{\bibfnamefont{G.~E.} \bibnamefont{Addison}},
  \bibinfo{author}{\bibfnamefont{P.~A.~R.} \bibnamefont{Ade}},
  \bibinfo{author}{\bibfnamefont{S.}~\bibnamefont{Aiola}},
  \bibinfo{author}{\bibfnamefont{D.}~\bibnamefont{Alonso}},
  \bibnamefont{et~al.}, \bibinfo{journal}{Journal of Cosmology and
  Astroparticle Physics} \textbf{\bibinfo{volume}{2020}}, \bibinfo{pages}{045}
  (\bibinfo{year}{2020}),
  \urlprefix\url{https://doi.org/10.1088/1475-7516/2020/12/045}.

\bibitem[{\citenamefont{{Planck Collaboration XI}}(2016)}]{Planck2015:ps}
\bibinfo{author}{\bibnamefont{{Planck Collaboration XI}}},
  \bibinfo{journal}{A\&A} \textbf{\bibinfo{volume}{594}}, \bibinfo{pages}{A11}
  (\bibinfo{year}{2016}),
  \urlprefix\url{https://doi.org/10.1051/0004-6361/201526926}.

\bibitem[{\citenamefont{{Addison} et~al.}(2016)\citenamefont{{Addison},
  {Huang}, {Watts}, {Bennett}, {Halpern}, {Hinshaw}, and
  {Weiland}}}]{Addison2016}
\bibinfo{author}{\bibfnamefont{G.~E.} \bibnamefont{{Addison}}},
  \bibinfo{author}{\bibfnamefont{Y.}~\bibnamefont{{Huang}}},
  \bibinfo{author}{\bibfnamefont{D.~J.} \bibnamefont{{Watts}}},
  \bibinfo{author}{\bibfnamefont{C.~L.} \bibnamefont{{Bennett}}},
  \bibinfo{author}{\bibfnamefont{M.}~\bibnamefont{{Halpern}}},
  \bibinfo{author}{\bibfnamefont{G.}~\bibnamefont{{Hinshaw}}},
  \bibnamefont{and} \bibinfo{author}{\bibfnamefont{J.~L.}
  \bibnamefont{{Weiland}}}, \bibinfo{journal}{\apj}
  \textbf{\bibinfo{volume}{818}}, \bibinfo{eid}{132} (\bibinfo{year}{2016}),
  \eprint{1511.00055}.

\bibitem[{\citenamefont{{Huang} et~al.}(2018)\citenamefont{{Huang}, {Addison},
  {Weiland}, and {Bennett}}}]{Huang2018}
\bibinfo{author}{\bibfnamefont{Y.}~\bibnamefont{{Huang}}},
  \bibinfo{author}{\bibfnamefont{G.~E.} \bibnamefont{{Addison}}},
  \bibinfo{author}{\bibfnamefont{J.~L.} \bibnamefont{{Weiland}}},
  \bibnamefont{and} \bibinfo{author}{\bibfnamefont{C.~L.}
  \bibnamefont{{Bennett}}}, \bibinfo{journal}{\apj}
  \textbf{\bibinfo{volume}{869}}, \bibinfo{eid}{38} (\bibinfo{year}{2018}),
  \eprint{1804.05428}.

\bibitem[{\citenamefont{Louis et~al.}(2019)\citenamefont{Louis, Li, and
  Tristram}}]{Louis19}
\bibinfo{author}{\bibfnamefont{T.}~\bibnamefont{Louis}},
  \bibinfo{author}{\bibfnamefont{Z.}~\bibnamefont{Li}}, \bibnamefont{and}
  \bibinfo{author}{\bibfnamefont{M.}~\bibnamefont{Tristram}},
  \bibinfo{journal}{Phys. Rev. D} \textbf{\bibinfo{volume}{100}},
  \bibinfo{pages}{103534} (\bibinfo{year}{2019}),
  \urlprefix\url{https://link.aps.org/doi/10.1103/PhysRevD.100.103534}.

\bibitem[{\citenamefont{{Galli} et~al.}(2022)\citenamefont{{Galli}, {Pogosian},
  {Jedamzik}, and {Balkenhol}}}]{Galli2021}
\bibinfo{author}{\bibfnamefont{S.}~\bibnamefont{{Galli}}},
  \bibinfo{author}{\bibfnamefont{L.}~\bibnamefont{{Pogosian}}},
  \bibinfo{author}{\bibfnamefont{K.}~\bibnamefont{{Jedamzik}}},
  \bibnamefont{and}
  \bibinfo{author}{\bibfnamefont{L.}~\bibnamefont{{Balkenhol}}},
  \bibinfo{journal}{\prd} \textbf{\bibinfo{volume}{105}}, \bibinfo{eid}{023513}
  (\bibinfo{year}{2022}), \eprint{2109.03816}.

\bibitem[{\citenamefont{La~Posta et~al.}(2021)\citenamefont{La~Posta, Louis,
  Garrido, Tristram, and Henrot-Versill\'e}}]{LaPosta2021}
\bibinfo{author}{\bibfnamefont{A.}~\bibnamefont{La~Posta}},
  \bibinfo{author}{\bibfnamefont{T.}~\bibnamefont{Louis}},
  \bibinfo{author}{\bibfnamefont{X.}~\bibnamefont{Garrido}},
  \bibinfo{author}{\bibfnamefont{M.}~\bibnamefont{Tristram}}, \bibnamefont{and}
  \bibinfo{author}{\bibfnamefont{S.}~\bibnamefont{Henrot-Versill\'e}},
  \bibinfo{journal}{Phys. Rev. D} \textbf{\bibinfo{volume}{104}},
  \bibinfo{pages}{023527} (\bibinfo{year}{2021}),
  \urlprefix\url{https://link.aps.org/doi/10.1103/PhysRevD.104.023527}.

\bibitem[{\citenamefont{{Ade} et~al.}(2019)\citenamefont{{Ade}, {Aguirre},
  {Ahmed}, {Aiola}, {Ali}, {Alonso}, {Alvarez}, {Arnold}, {Ashton},
  {Austermann} et~al.}}]{SOoverview}
\bibinfo{author}{\bibfnamefont{P.}~\bibnamefont{{Ade}}},
  \bibinfo{author}{\bibfnamefont{J.}~\bibnamefont{{Aguirre}}},
  \bibinfo{author}{\bibfnamefont{Z.}~\bibnamefont{{Ahmed}}},
  \bibinfo{author}{\bibfnamefont{S.}~\bibnamefont{{Aiola}}},
  \bibinfo{author}{\bibfnamefont{A.}~\bibnamefont{{Ali}}},
  \bibinfo{author}{\bibfnamefont{D.}~\bibnamefont{{Alonso}}},
  \bibinfo{author}{\bibfnamefont{M.~A.} \bibnamefont{{Alvarez}}},
  \bibinfo{author}{\bibfnamefont{K.}~\bibnamefont{{Arnold}}},
  \bibinfo{author}{\bibfnamefont{P.}~\bibnamefont{{Ashton}}},
  \bibinfo{author}{\bibfnamefont{J.}~\bibnamefont{{Austermann}}},
  \bibnamefont{et~al.}, \bibinfo{journal}{\jcap}
  \textbf{\bibinfo{volume}{2019}}, \bibinfo{eid}{056} (\bibinfo{year}{2019}),
  \eprint{1808.07445}.

\bibitem[{\citenamefont{{Abazajian} et~al.}(2019)\citenamefont{{Abazajian},
  {Addison}, {Adshead}, {Ahmed}, {Allen}, {Alonso}, {Alvarez}, {Anderson},
  {Arnold}, {Baccigalupi} et~al.}}]{CMB-S4}
\bibinfo{author}{\bibfnamefont{K.}~\bibnamefont{{Abazajian}}},
  \bibinfo{author}{\bibfnamefont{G.}~\bibnamefont{{Addison}}},
  \bibinfo{author}{\bibfnamefont{P.}~\bibnamefont{{Adshead}}},
  \bibinfo{author}{\bibfnamefont{Z.}~\bibnamefont{{Ahmed}}},
  \bibinfo{author}{\bibfnamefont{S.~W.} \bibnamefont{{Allen}}},
  \bibinfo{author}{\bibfnamefont{D.}~\bibnamefont{{Alonso}}},
  \bibinfo{author}{\bibfnamefont{M.}~\bibnamefont{{Alvarez}}},
  \bibinfo{author}{\bibfnamefont{A.}~\bibnamefont{{Anderson}}},
  \bibinfo{author}{\bibfnamefont{K.~S.} \bibnamefont{{Arnold}}},
  \bibinfo{author}{\bibfnamefont{C.}~\bibnamefont{{Baccigalupi}}},
  \bibnamefont{et~al.}, \bibinfo{journal}{arXiv e-prints}
  \bibinfo{eid}{arXiv:1907.04473} (\bibinfo{year}{2019}), \eprint{1907.04473}.

\bibitem[{\citenamefont{{Torrado} and {Lewis}}(2021)}]{Cobaya}
\bibinfo{author}{\bibfnamefont{J.}~\bibnamefont{{Torrado}}} \bibnamefont{and}
  \bibinfo{author}{\bibfnamefont{A.}~\bibnamefont{{Lewis}}},
  \bibinfo{journal}{\jcap} \textbf{\bibinfo{volume}{2021}}, \bibinfo{eid}{057}
  (\bibinfo{year}{2021}), \eprint{2005.05290}.

\bibitem[{\citenamefont{{Torrado} and {Lewis}}(2019)}]{cobayaascii}
\bibinfo{author}{\bibfnamefont{J.}~\bibnamefont{{Torrado}}} \bibnamefont{and}
  \bibinfo{author}{\bibfnamefont{A.}~\bibnamefont{{Lewis}}},
  \emph{\bibinfo{title}{{Cobaya: Bayesian analysis in cosmology}}}
  (\bibinfo{year}{2019}), \eprint{1910.019}.

\bibitem[{\citenamefont{{Planck Collaboration}
  et~al.}(2016)\citenamefont{{Planck Collaboration}, {Ade}, {Aghanim},
  {Arnaud}, {Ashdown}, {Aumont}, {Baccigalupi}, {Banday}, {Barreiro},
  {Bartlett} et~al.}}]{Planck2015:cosmo}
\bibinfo{author}{\bibnamefont{{Planck Collaboration}}},
  \bibinfo{author}{\bibfnamefont{P.~A.~R.} \bibnamefont{{Ade}}},
  \bibinfo{author}{\bibfnamefont{N.}~\bibnamefont{{Aghanim}}},
  \bibinfo{author}{\bibfnamefont{M.}~\bibnamefont{{Arnaud}}},
  \bibinfo{author}{\bibfnamefont{M.}~\bibnamefont{{Ashdown}}},
  \bibinfo{author}{\bibfnamefont{J.}~\bibnamefont{{Aumont}}},
  \bibinfo{author}{\bibfnamefont{C.}~\bibnamefont{{Baccigalupi}}},
  \bibinfo{author}{\bibfnamefont{A.~J.} \bibnamefont{{Banday}}},
  \bibinfo{author}{\bibfnamefont{R.~B.} \bibnamefont{{Barreiro}}},
  \bibinfo{author}{\bibfnamefont{J.~G.} \bibnamefont{{Bartlett}}},
  \bibnamefont{et~al.}, \bibinfo{journal}{\aap} \textbf{\bibinfo{volume}{594}},
  \bibinfo{eid}{A13} (\bibinfo{year}{2016}), \eprint{1502.01589}.

\bibitem[{\citenamefont{Aghanim et~al.}(2020)}]{Planck2018:like}
\bibinfo{author}{\bibfnamefont{N.}~\bibnamefont{Aghanim}} \bibnamefont{et~al.}
  (\bibinfo{collaboration}{Planck}), \bibinfo{journal}{Astron. Astrophys.}
  \textbf{\bibinfo{volume}{641}}, \bibinfo{pages}{A5} (\bibinfo{year}{2020}),
  \eprint{1907.12875}.

\bibitem[{\citenamefont{Hill et~al.}(2021)}]{ACTede}
\bibinfo{author}{\bibfnamefont{J.~C.} \bibnamefont{Hill}} \bibnamefont{et~al.}
  (\bibinfo{year}{2021}), \eprint{2109.04451}.

\bibitem[{\citenamefont{Lewis}(2019)}]{getdist}
\bibinfo{author}{\bibfnamefont{A.}~\bibnamefont{Lewis}} (\bibinfo{year}{2019}),
  \eprint{1910.13970}, \urlprefix\url{https://getdist.readthedocs.io}.

\bibitem[{\citenamefont{{Cartis}
  et~al.}(2018{\natexlab{a}})\citenamefont{{Cartis}, {Fiala}, {Marteau}, and
  {Roberts}}}]{Cartis2018a}
\bibinfo{author}{\bibfnamefont{C.}~\bibnamefont{{Cartis}}},
  \bibinfo{author}{\bibfnamefont{J.}~\bibnamefont{{Fiala}}},
  \bibinfo{author}{\bibfnamefont{B.}~\bibnamefont{{Marteau}}},
  \bibnamefont{and}
  \bibinfo{author}{\bibfnamefont{L.}~\bibnamefont{{Roberts}}},
  \bibinfo{journal}{arXiv e-prints} \bibinfo{eid}{arXiv:1804.00154}
  (\bibinfo{year}{2018}{\natexlab{a}}), \eprint{1804.00154}.

\bibitem[{\citenamefont{{Cartis}
  et~al.}(2018{\natexlab{b}})\citenamefont{{Cartis}, {Roberts}, and
  {Sheridan-Methven}}}]{Cartis2018b}
\bibinfo{author}{\bibfnamefont{C.}~\bibnamefont{{Cartis}}},
  \bibinfo{author}{\bibfnamefont{L.}~\bibnamefont{{Roberts}}},
  \bibnamefont{and}
  \bibinfo{author}{\bibfnamefont{O.}~\bibnamefont{{Sheridan-Methven}}},
  \bibinfo{journal}{arXiv e-prints} \bibinfo{eid}{arXiv:1812.11343}
  (\bibinfo{year}{2018}{\natexlab{b}}), \eprint{1812.11343}.

\bibitem[{\citenamefont{Lewis et~al.}(2000)\citenamefont{Lewis, Challinor, and
  Lasenby}}]{Lewis2000}
\bibinfo{author}{\bibfnamefont{A.}~\bibnamefont{Lewis}},
  \bibinfo{author}{\bibfnamefont{A.}~\bibnamefont{Challinor}},
  \bibnamefont{and} \bibinfo{author}{\bibfnamefont{A.}~\bibnamefont{Lasenby}},
  \bibinfo{journal}{The Astrophysical Journal} \textbf{\bibinfo{volume}{538}},
  \bibinfo{pages}{473} (\bibinfo{year}{2000}),
  \urlprefix\url{https://doi.org/10.1086/309179}.

\bibitem[{\citenamefont{Howlett et~al.}(2012)\citenamefont{Howlett, Lewis,
  Hall, and Challinor}}]{Howlett2012}
\bibinfo{author}{\bibfnamefont{C.}~\bibnamefont{Howlett}},
  \bibinfo{author}{\bibfnamefont{A.}~\bibnamefont{Lewis}},
  \bibinfo{author}{\bibfnamefont{A.}~\bibnamefont{Hall}}, \bibnamefont{and}
  \bibinfo{author}{\bibfnamefont{A.}~\bibnamefont{Challinor}},
  \bibinfo{journal}{Journal of Cosmology and Astroparticle Physics}
  \textbf{\bibinfo{volume}{2012}}, \bibinfo{pages}{027} (\bibinfo{year}{2012}),
  \urlprefix\url{https://doi.org/10.1088/1475-7516/2012/04/027}.

\end{thebibliography}

\end{document}